\documentclass[aps,prb,floatfix,twocolumn,superscriptaddress,showpacs]{revtex4-1}
\usepackage{setspace}
\usepackage{graphicx}
\usepackage{epstopdf}

\usepackage[colorlinks=true,
            linkcolor=red,
            urlcolor=blue,
            citecolor=blue]{hyperref}
            
\usepackage{color}

\DeclareTextCompositeCommand{\'}{OT1}{i}{\@tabacckludge'\i}

\graphicspath{{images/}}

\newcommand{\simlt}
      {\ifmmode       { \raisebox{-.8em}{$<$}\atop\sim}
         \else        {$\raisebox{-.8em}{$<$}\atop\sim$}
      \fi}

\begin{document}
\title{Observation of a well-defined hybridization gap and in-gap states \\ on the SmB$_6$ (001) surface}
\author{Zhixiang Sun}
\email{z.sun@ifw-dresden.de}
\affiliation{Max-Planck-Institut f\"ur Festk\"orperforschung, Heisenbergstr. 1, D-70569 Stuttgart, Germany}
\author{Ana Maldonado}
\affiliation{Max-Planck-Institut f\"ur Festk\"orperforschung, Heisenbergstr. 1, D-70569 Stuttgart, Germany}
\affiliation{SUPA, School of Physics and Astronomy, University of St. Andrews, North Haugh, St. Andrews, Fife, KY16 9SS, United Kingdom}
\author{Wendel S. Paz}
\affiliation{Departamento de Fisica de la Materia Condensada, Universidad Aut$\acute{o}$noma de Madrid, Cantoblanco, 28049 Madrid, Spain}
\author{Dmytro S. Inosov}
\affiliation{Institut f\"ur Festk\"orper- und Materialphysik, Technische Universit\"at Dresden, D-01062, Dresden, Germany  }
\affiliation{Max-Planck-Institut f\"ur Festk\"orperforschung, Heisenbergstr. 1, D-70569 Stuttgart, Germany}
\author{Andreas P. Schnyder}
\affiliation{Max-Planck-Institut f\"ur Festk\"orperforschung, Heisenbergstr. 1, D-70569 Stuttgart, Germany}
\author{J. J. Palacios}
\affiliation{Departamento de Fisica de la Materia Condensada, Universidad Aut$\acute{o}$noma de Madrid, Cantoblanco, 28049 Madrid, Spain}
\affiliation{Condensed Matter Physics Center (IFIMAC), Universidad Aut$\acute{o}$noma de Madrid, Cantoblanco, 28049 Madrid, Spain}
\affiliation{Instituto Nicol\'as Cabrera (INC), Universidad Aut$\acute{o}$noma de Madrid, Cantoblanco, 28049 Madrid, Spain}
\author{Natalya Yu. Shitsevalova}
\affiliation{Frantsevich Institute for Problems of Material Sciences of NASU, 3 Krzhyzhanovsky str., 03680 Kiev, Ukraine}
\author{Vladimir B. Filipov}
\affiliation{Frantsevich Institute for Problems of Material Sciences of NASU, 3 Krzhyzhanovsky str., 03680 Kiev, Ukraine}
\author{Peter Wahl}
\email{wahl@st-andrews.ac.uk}
\affiliation{SUPA, School of Physics and Astronomy, University of St. Andrews, North Haugh, St. Andrews, Fife, KY16 9SS, United Kingdom}
\affiliation{Max-Planck-Institut f\"ur Festk\"orperforschung, Heisenbergstr. 1, D-70569 Stuttgart, Germany}


\begin{abstract}
The rise of topology in condensed matter physics has generated strong interest in identifying novel quantum materials in which topological protection is driven by electronic correlations. Samarium hexaboride is a Kondo insulator for which it has been proposed that a band inversion between $5d$ and $4f$ bands gives rise to topologically protected surface states. However, unambiguous proof of the existence and topological nature of these surface states is still missing, and its low-energy electronic structure is still not fully established. Here we present a study of samarium hexaboride by ultra-low-temperature scanning tunneling microscopy and spectroscopy. We obtain clear atomically resolved topographic images of the sample surface. Our tunneling spectra reveal signatures of a hybridization gap with a size of about $8\ \mathrm{meV}$ and with a reduction of the differential conductance inside the gap by almost half, and surprisingly, several strong resonances below the Fermi level. The spatial variations of the energy of the resonances point towards a microscopic variation of the electronic states by the different surface terminations. High-resolution tunneling spectra acquired at $100\ \mathrm{mK}$ reveal a splitting of the Kondo resonance, possibly due to the crystal electric field. 
\end{abstract}

\pacs{71.27.+a, 71.28.+d, 73.20.-r}
\maketitle

\section{Introduction}
Samarium hexaboride is a Kondo lattice compound in which a hybridization gap at the Fermi level is formed below a characteristic temperature $T^\star$ due to the Kondo screening effect \cite{menth1969magnetic, nickerson1971physical}. It was also the first mixed valence compound that has been established, the valence of samarium fluctuates between Sm$^{2+}$(4$f^6$) and Sm$^{3+}$(4$f^5$) with an average value of about +2.6--2.7 at ambient conditions \cite{vainshtein1965, cohen1970electronic, chazalviel1976study}. In the resistivity measurements, below a characteristic temperature $T^\star\sim 50\ \mathrm{K}$, an exponential increase in the resistivity is observed with the lowering temperature, which is the typical behavior for a metal-to-semiconductor transition \cite{frantzeskakis2013kondo}. This is attributed to the opening of a Kondo hybridization gap \cite{Riseborough2000}. Though the size of the gap should be of the same order of magnitude as the temperature $2k_\mathrm{B}T^\star$ of the crossover, depending on measurements, the reported size of the hybridization gap varies between 3 and 20 meV \cite{nanba1993gap, takigawa1981nmr, zhang2013hybridization}. Furthermore, multi-gap features have also been reported from optical spectroscopy \cite{gorshunov1999low, yamaguchi2013different}.

Another puzzle is the observation of the resistivity saturating at temperatures below $T_h\sim 5\mathrm{\ K}$, rather than rising further \cite{gabani2015surface}. The behavior is attributed to an additional conductance channel, for which conduction through topologically protected surface states is one possible interpretation. There have been a number of alternative theoretical proposals to explain this observation, e.g., formation of a Wigner lattice \cite{kasuya1979valence}, Mott minimum conductivity \cite{allen1979large}, phonon bound states due to magnetoelastic coupling \cite{nyhus1997low}, impurity bands \cite{allen1979large} and trivial surface states \cite{hlawenka2018samarium}. Clear experimental evidence for the origin of the in-gap states is still missing. Recent theoretical calculations show that SmB$_6$ is a promising candidate as a topological Kondo insulator \cite{dzero2010topological, dzero2012theory, lu2013correlated, takimoto2011smb6, alexandrov2013cubic, ye2013topological}. Within this interpretation, the Kondo hybridization drives a band inversion in the band structure of SmB$_6$ and leads to a hybridization gap in which topologically non-trivial surface states are stabilized \cite{RevModPhys.83.1057, hasan2010colloquium}. A wide range of methods, such as angle-resolved photoemission spectroscopy (ARPES) \cite{frantzeskakis2013kondo, neupane2013surface, zhu2013polarity, xu2013surface, suga2014spin}, electronic transport \cite{wolgast2013low, zhang2013hybridization}, x-ray reflectometry \cite{zabolotnyy2018chemical}, and scanning probe methods \cite{yee2013imaging, rossler2014hybridization, ruan2014emergence} have been applied in an attempt to establish the surface electronic structure and search for evidence of the topologically protected states. While there is evidence for surface states, their topological nature remains ambiguous.

Here we study the low temperature electronic properties of an unreconstructed (001) surface of SmB$_6$ by ultra-low temperature scanning tunneling microscopy and spectroscopy (STM/STS). Our results show a hybridization gap of about 8 meV at $E_\mathrm{F}$, with a strong Kondo-like resonance. The tunneling spectra show distinct resonance states around the Fermi level, which we attribute to the samarium $4f$-states and which develop at very low temperatures a fine structure consisting of a series of multiple peaks. The spatial variations of the resonance show evidence for local doping, moving them away from the Fermi level near surface defects. To understand the surface termination, we compare our data to density functional theory (DFT) calculations of the surface morphology as well as its work function and energetics. Our spectroscopic observations are consistent with the existence of in-gap surface states. 

\begin{figure*}
\centering
\includegraphics[width=.95\textwidth]{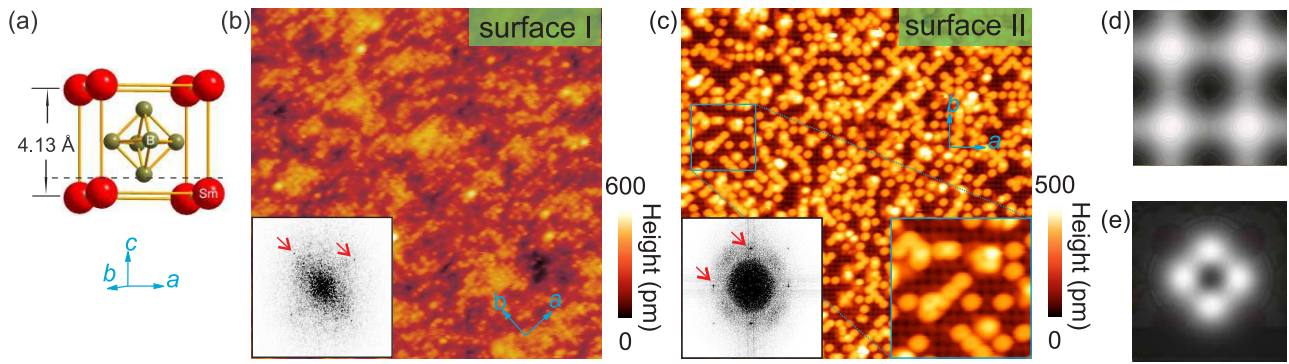}
\caption{Crystal structure and topographic imaging. (a) Illustration of the crystal structure of SmB$_6$. (b) Topographic STM image of the sample surface I, $20\times 20\ \mathrm{nm}^2$ ($V_b=100\ \mathrm{mV}$, $I=1.0\ \mathrm{nA}$). Inset: Fourier transformation of the topography, showing the peaks associated with the square lattice (marked by red arrows). (c) Topographic STM image of the sample \textcolor{red}{surface II}, $20\times 20\ \mathrm{nm}^2$ ($V_b=200\ \mathrm{mV}$, $I=50\ \mathrm{pA}$). Inset at bottom left: Fourier transformation of the topography, besides the peaks associated with the square lattice (marked by red arrows) higher order peaks can be seen. Bottom right inset: zoom in topography at atomic scale which shows the atomically resolved surface and the fine structure of the surface clusters ($4\times4\ \mathrm{nm}^2$). Note that topography shown in (c) is rotated $45^\circ$ with regards to panels (b). (d) Simulated topography of the Sm terminated surface. (e) Simulated topography for a $\mathrm{B}_5$ cluster on the Sm terminated surface.}
\label{topos}
\end{figure*}

\section{Methods}
\subsection*{Scanning Tunneling Microscopy/Spectroscopy}
STM experiments were performed in a home-built low temperature STM, operating at temperatures down to $10\ \mathrm{mK}$ \cite{Singh2013} in cryogenic vacuum. Samples are prepared by {\it in-situ} cleaving. We used STM tips cut from a $0.2\ \mathrm{mm}$ PtIr wire. Bias voltages are applied to the sample, with the tip at virtual ground. Differential conductance spectra have been recorded through a lock-in amplifier with a frequency of 411 Hz. The $\mathrm{SmB}_6$ sample is loaded into the STM chamber from a load-lock at a pressure of about $10^{-6}$ mbar, and cleaved in cryogenic vacuum at a cleaving stage at 4K perpendicular to the $(001)$ axis at temperatures below $20\ \mathrm{K}$. The sample growth procedure is similar to the growth of $\mathrm{CeB}_6$ as described in Ref.~\onlinecite{friemel2012resonant}.

\subsection*{Calculations}
Our calculations were performed based on the framework of DFT, as implemented in the Quantum ESPRESSO package \cite{giannozzi_quantum_2009}. The Perdew-Burke-Ernzerhof generalized gradient approximation (GGA-PBE) \cite{perdew_generalized_1996} was adopted for the exchange-correlation functional.  The electron-ion interaction is described using the norm-conserving Troullier-Martins pseudopotentials \cite{troullier_efficient_1991}.  The energy cut-off for the plane-wave basis set is taken to be $120\mathrm{Ry}$ with a charge density cut-off of $500\mathrm{Ry}$. We have used a Monkhorst-Pack \cite{monkhorst_special_1976} scheme with a $k$-mesh for the Brillouin zone integration for the supercells with one unit cell terminated with samarium, hexaboride ($\mathrm B_6$), pentaboride ($\mathrm{B}_5$) and boron ($\mathrm B_1$) surfaces and a mesh for the supercells formed by three unit cells yielding surfaces terminated with clusters of $\mathrm B_6$ and $\mathrm B_5$ separated by $11.0\ \mathrm\AA$.  In all calculations the lattice parameter was kept fixed at the experimental value $a=4.13\ \mathrm\AA$, and we used $15\ \mathrm\AA$ of vacuum to minimize interactions between the surfaces of the slabs.

The electrostatic potential average is calculated from the electronic density $n(\mathbf r)$. The plane-averaged electronic density is defined,
\begin{equation}
\bar{n}=\frac{1}{S}\int_S n(\mathbf r)\mathrm d x\mathrm d y,
\end{equation}
where the $z$-axis is perpendicular to the slab surface $S$. The electrostatic potential $V(\mathbf r)$ is related to the total charge density, including ionic charge, via the Poisson equation.

For the simulated STM images we used the Tersoff-Hamann theory \cite{tersoff_theory_1985}, with a voltage between the sample and the tip of 1.0 V for the unoccupied states.

\section{Results}
\subsection*{Surface topography}
SmB$_6$ has a CsCl-like crystal structure as shown in Fig.~\ref{topos}(a). Crystals have been oriented in the (001) direction prior to cleavage. The material does not exhibit a strongly preferred natural cleavage plane, thus different terminations can be expected to occur and have been reported previously \cite{yee2013imaging, rossler2014hybridization, koitzsch2016nesting}. We have observed two types of surfaces. The first (surface I), shown in Fig.~\ref{topos}(b), is flat on the scale of a few Angstroms and exhibits substantial inhomogeneity. Atomic structure with very short range regularity can be seen on the surface. Nevertheless, in the Fourier transformation we can still observe atomic peaks consistent with the lattice constant of $\mathrm{SmB}_6$ (see the inset of Fig. \ref{topos}(b)). The second surface type we have observed (surface II) is shown in Fig.~\ref{topos}(c), it exhibits nanometer sized areas which show a clean and flat atomic lattice (see inset). On surface II, identical clusters can be observed, which cover the surface homogenously on the macroscopic scale. It is also noticed that the surface clusters are much larger in size than the atomic protrusions on surface I. On this surface, the majority of surface clusters have a squarish ring-like appearance under certain bias. Similar ring-like defects have been reported previously and interpreted as boron clusters \cite{ruan2014emergence}. We are interested in a better understanding of the electronic effects of these squarish structures. Because of the inhomogeneity of surface I, we cannot uniquely identify what termination it corresponds to, and therefore concentrate in the following discussion on surface II.

\begin{figure*}
\begin{center}
\includegraphics[width=0.9\textwidth]{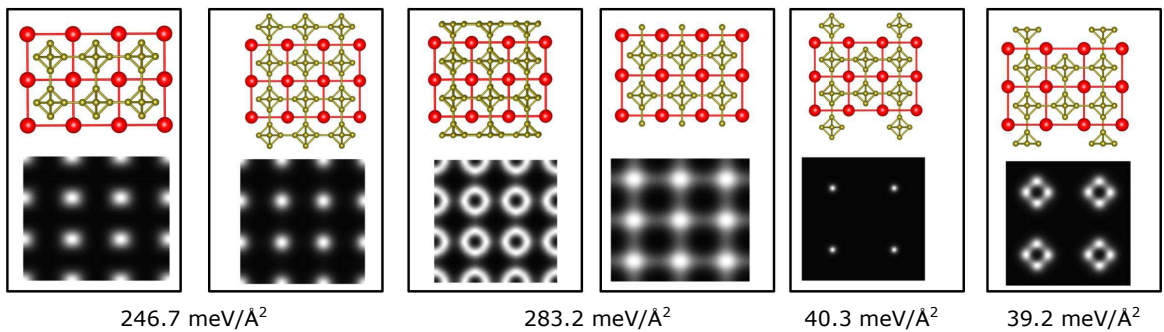}
\vspace{0cm}
\end{center}
\caption{Comparison of different surface terminations, simulated STM topographies and estimated surface energies $\gamma$ through Eq. (\ref{surfaceenergy}) from DFT calculations. From left to right: Sm-surface; hexaboride-surface; pentaboride-surface; boron-surface; hexaboride-cluster and pentaboride-cluster.}
\label{suppl}
\end{figure*}

From comparison with DFT calculations, we can identify the surface clusters as disrupted boron cages on a samarium-terminated surface, supported by the DFT simulated topography (as shown in Fig.~\ref{topos}(d) and (e)). Simulations for other surface terminations are shown in Fig.~\ref{suppl}.\\

\begin{figure}
\begin{center}
\includegraphics[width=.45\textwidth]{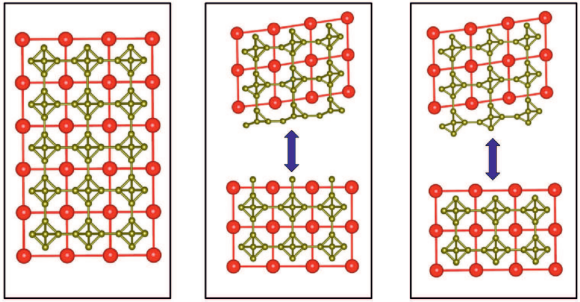}
\vspace{0cm}
\end{center}
\caption{Schematic process for the creation of two surfaces (inequivalent in this case) Sm-B$_6$ and B$_1$-B$_5$.  Note that in the two right panels the upper part of the crystal is shown distorted to highlight that it is cleaved of the lower part.}
\label{creation}
\end{figure}

\begin{figure}
\begin{center}
\includegraphics[width=.48\textwidth]{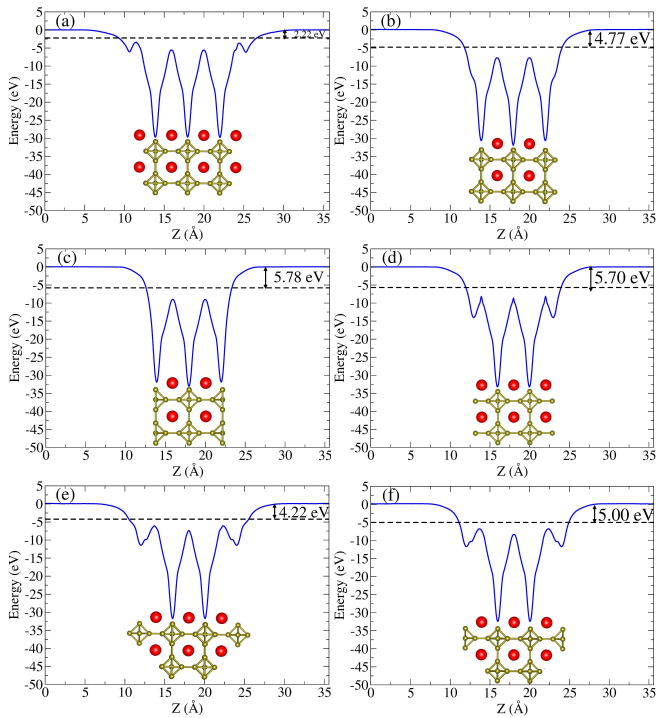}
\vspace{0cm}
\end{center}
\caption{Electrostatic potential of the SmB$_6$ surface between the slab and vacuum, for surfaces terminated at (a) Sm-atom, (b) hexaboride, (c) pentaboride, (d) boron-atom, (e) hexaboride-cluster, and (f) pentaboride-cluster.}
\label{cutfigure}
\end{figure}

In order to identify the most likely surface terminations from theory, we have calculated the surface formation energy for different types of surfaces. The surface energy is defined here as the energy required to create a new surface starting from a bulk system or, in other words, the energy required to break a bulk sample, creating two surfaces. Figure \ref{creation} schematically shows the formation process (the upper slab is shown distorted for clarity). In our calculations the surface energy can thus be determined by taking the energy difference between the total energy of two slabs (formed after cutting a bulk sample) and an equivalent bulk reference. Using a supercell model for each slab, the surface energy $\gamma$ at $T = 0\ \mathrm K$ of a clean surface is given by

\begin{equation}
\gamma=\frac{1}{2A} \left(E_{\mathrm{slab}}^\mathrm{total}-E_\mathrm{bulk}^\mathrm{ref}\right),
\label{surfaceenergy}
\end{equation}

where $E_{\mathrm{slab}}^\mathrm{total}$ and $E_\mathrm{bulk}^\mathrm{ref}$ are the total energies given by the sum of two separate slabs and of the bulk reference, respectively (see Fig. \ref{cutfigure}). $A$ is the surface unit area, and the factor 1/2 appears because $E_\mathrm{slab}^\mathrm{total}$ contains two surfaces. Since these two surfaces are not necessarily equivalent, the surface energy $\gamma$ represents the mean value of the two (different) surfaces. The pairs of surfaces formed after the cut are: Firstly, Sm-B$_6$ and B$_1$-B$_5$ with both surfaces fully covered with boron clusters (or uncovered) and, second, partially covered (cluster-type surface). In the latter cases the two surfaces end up sharing the B$_6$ and B$_5$-B$_1$ clusters, and the energies have been calculated for a 33\%-66\% coverage. These turn out to be the lowest in energy with a small advantage for the B$_5$-B$_1$ cluster surface. This is consistent with the experimental observations.

\begin{figure}
\begin{center}
\includegraphics[width=.45\textwidth]{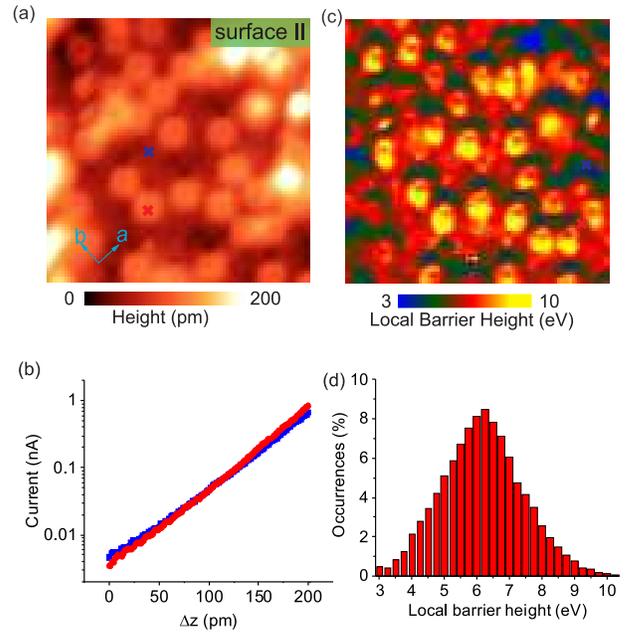}
\end{center}
\caption{Determination of surface termination. (a) Topographic STM images acquired simultaneously with a map of the local barrier height extracted from $I(z)$ curves ($6.7\times6.7\ \mathrm{nm}^2$, $V_b=300\ \mathrm{mV}, I=50\ \mathrm{pA}$). (b) Two typical $I(z)$ curves obtained on a clean spot on the surface (marked by blue cross in (a)) and on top of a cluster (marked by a red cross in (a)). The two curves (plotted on a logarithmic axis) exhibit significantly different local barrier heights, with a substantially higher one on top of the cluster. (c) Spatial map of the local barrier height acquired simultaneously with the topography shown in (a), positions of the two traces shown in (b) are marked by crosses. (d) Histogram of the local barrier height of the map shown in (c). }
\label{workfunction}
\end{figure}

We have measured the local barrier height in surface II, as shown in Fig.~\ref{workfunction}(a). A difference in work function between Sm atoms and $\mathrm{B}_6$ octahedra or boron clusters can be expected to lead to variations in the local barrier height. The local barrier height can be obtained from measurements of the tunneling current as a function of tip-sample distance, $I(z)$. Fig.~\ref{workfunction}(b) shows two examples of $I(z)$ curves obtained on one of the defects and on a clean patch of the surface (as indicated in Fig.~\ref{workfunction}(a)). The local barrier height is obtained from a fit of an exponential decay to these curves, revealing a substantially lower barrier height on the defect-free areas compared to the defects. A spatial map of the barrier height is shown in Fig.~\ref{workfunction}(c), obtained simultaneously with the topographic image shown in Fig.~\ref{workfunction}(a). The spatial maps show that the patches of the clean atomically resolved surface have a local barrier height on the order of $4\ \mathrm{eV}$, whereas on defects, a substantially larger local barrier height of $7\ \mathrm{eV}$ is found. This behaviour indicates that the defects have a large electron affinity compared to the clean surface, consistent with the interpretation of these defects as Boron atoms or clusters. Samarium adatoms would rather be expected to lead to a local decrease in the barrier height, due to their charge and the Smoluchowski effect \cite{smoluchowski_anisotropy_1941}. The assignment is consistent with calculations of the work function for different surface terminations which show a very low work function on the order of $2\ \mathrm{eV}$ for the samarium terminated surface, whereas $\mathrm B$ terminated surfaces have a work function at least twice as high. It should be noted that the experiment does not directly measure the work function (or a local equivalent of the work function), but the local barrier height between the tip and the sample, which is related to the work function. For the case of a clean surface and tip with work functions $\Phi_\mathrm s$ and $\Phi_\mathrm t$, respectively, the current will increase as $\exp(\kappa\Delta z$), with $\kappa=\sqrt{\frac{m_e}{2\hbar^2}(\Phi_\mathrm s+\Phi_\mathrm t)}$, for bias voltages $V<<\Phi_{\mathrm s, \mathrm t}$, and where $m_e$ is the electron mass.
The calculations also do confirm that surface terminations with clusters of $\mathrm{B}_5$ or $\mathrm{B}_6$ on a Sm terminated surface have much lower surface energies compared to clean B or Sm terminations (see Fig. \ref{cutfigure}).

\begin{figure}
\begin{center}
\includegraphics[width=.48\textwidth]{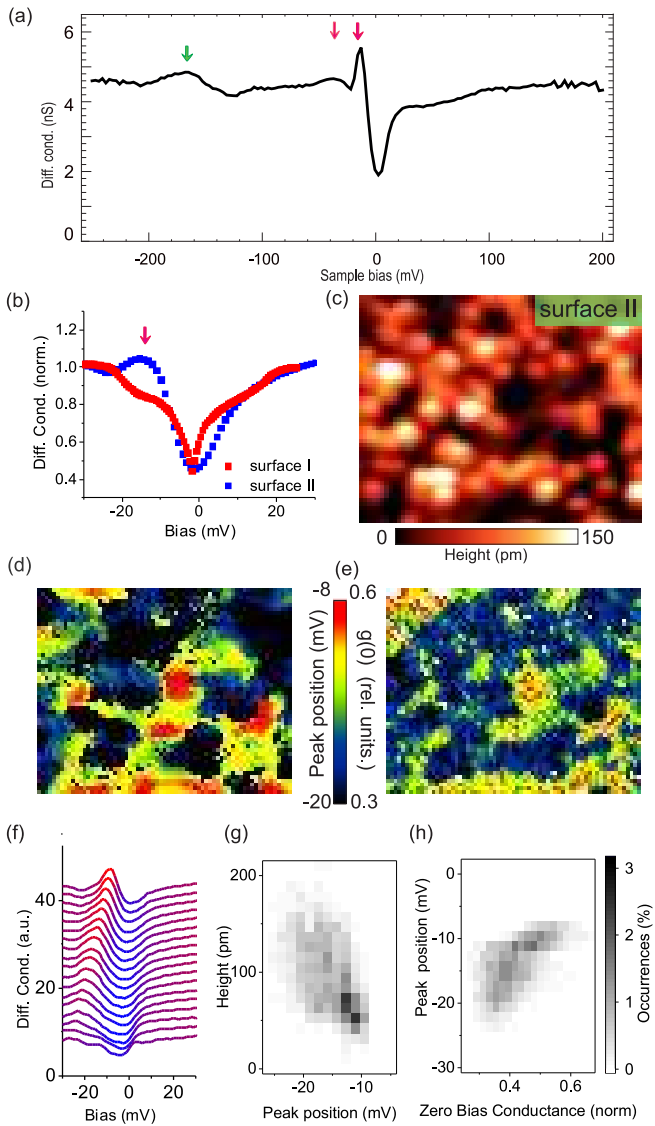}
\end{center}
\caption{Tunneling conductance spectroscopy. (a) A typical large-range tunneling
conductance spectrum taken on surface II. Spectrum setpoint: $V_b$ = 200 meV, $I$ = 0.9 nA, with a modulation voltage $V_\mathrm{{mod}}$ = 0.5 mV. $T$ = 10 K. (b) Spatially averaged tunneling spectra obtained on the two types of surface terminations shown in Fig.~\ref{topos}. (c) Topography and (d), (e) spatial maps of the peak position and the differential conductance at zero-bias extracted from a spectroscopic map ($7.0\times5.6\ \mathrm{nm}^2$, $T=10\ \mathrm K$, $V_b=200\ \mathrm{mV}, I=50\ \mathrm{pA}$). (f) Averages over spectra which exhibit the same peak position, from closest to zero-bias (top) to furthest (bottom). (g) Correlation between peak position and topographic height confirms this behaviour of the surface impurities moving the peak to lower energies, with a correlation coefficient of $-0.5$. (h) Correlation between peak position and zero-bias conductance $g(V)$ (normalized by the conductance at $V=-50\ \mathrm{mV}$) shows that as the peak shifts away from the Fermi energy, the zero-bias conductance is suppressed, with a correlation coefficient of $0.5$.}
\label{spectroscopy}
\end{figure}

\subsection*{Spatial variation of the tunneling spectrum}
In order to probe the electronic structure, we have measured the differential tunneling conductance spectra (${\mathrm d}I(V_b)/{\mathrm d}V_b$), which is, under certain assumptions, proportional to the local density of states (LDOS). In Fig.~\ref{spectroscopy}(a) we show a large-range spectrum taken on surface II. From it we can see that the first sharp resonance peak below $E_\mathrm{F}$ is found around $-15\ \mathrm{mV}$ and a second one near $-40\ \mathrm{mV}$ as indicated by the red arrows. Further away from $E_\mathrm{F}$, at about $-180\ \mathrm{mV}$, there is an additional broader resonance. We can relate the positions of the peaks to the energies of some of the Sm $4f$ bands. Comparing with the $4f$ bands observed in previous ARPES measurements (e.g., Ref. \onlinecite{frantzeskakis2013kondo, zhu2013polarity}) and dynamical mean field theory (DMFT) calculations \cite{deng2013plutonium, denlinger2013temperature}, we can see a one-to-one correspondence between the peak positions and the $4f$ band energy positions probed by ARPES.

Typical spatially averaged conductance spectra for the two types of surfaces are shown in Fig.~\ref{spectroscopy}(b). On surface I, the spectrum exhibits two gap-like features, whereas for surface II, besides a gap-like feature close to the Fermi energy $E_\mathrm{F}$, we observe a peak below the Fermi level (around $-15$ meV, as indicated by the arrow in Fig.~\ref{spectroscopy}(b)). The overall shape and energy scales of the spectra taken at 10 K are consistent with previously reported tunneling spectra \cite{yee2013imaging, rossler2014hybridization}. The spectra obtained on surface II resemble Fano lineshapes, as observed frequently in heavy fermion compounds \cite{Aynajian2010, Aynajian2012, Park2012}.  To further clarify the effects of surface termination and the nature of the Fano resonance feature, we need to measure the electronic structure at a much lower temperature. In the remainder of this work we will concentrate on the surface shown in Fig.~\ref{topos}(c), surface II, unless stated otherwise. To elucidate the physics of this lineshape and its relation to the band structure of $\mathrm{SmB}_6$, we have analyzed the dependence of the spectra on the presence of boron adatoms. Figures~\ref{spectroscopy}(c), (d) and (e) show a topographic STM image as well as maps of the energy of the peak and of the zero-bias conductance. Direct comparison already indicates that the peak is closest to the Fermi energy, at $-8\ \mathrm{mV}$, for clean surface areas, whereas it is shifted to more negative energies close to or on the boron clusters. Averaging over spectra that exhibit the same energy of the peak, as shown in Fig.~\ref{spectroscopy}(f), shows that while the amplitude of the peak is directly correlated with its energy, the gap remains mostly independent of that. This indicates that the two have separate origins, rather than both emerging from the hybridization gap or a Fano lineshape. Figures~\ref{spectroscopy}(g, h) show two-dimensional histograms between the peak position and the topographic height, and the peak energy and the zero-bias conductance. They demonstrate that (i) the peak position is directly correlated with the local topographic height (see Fig.~\ref{spectroscopy}(g)), and (ii) that the peak position exhibits a strong correlation with the zero-bias conductance (see Fig.~\ref{spectroscopy}(h)). On top of boron adatoms, where the topographic height is larger, the peak position is shifted to more negative energy and hence away from the Fermi energy.

\subsection*{Spectroscopy at ultra-low temperatures}
\begin{figure}
\begin{center}
\includegraphics[width=.4\textwidth]{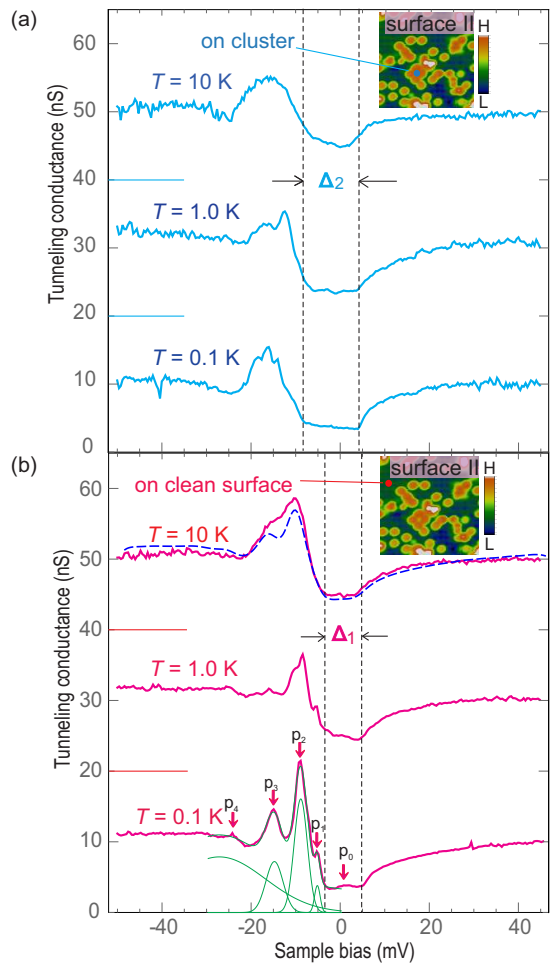}
\end{center}
\caption{ Ultra-low temperature spectra and the temperature dependence. Conductance spectra measured with high energy resolution at temperatures down to $0.1\ \mathrm{K}$ on surface II, shifted vertically by 20 nS for clarity. (a) The tunnelling spectra taken on the  boron clusters at 0.1 K, 1.0 K and 10.0 K. $\Delta_2$ is the label of the hybridization gap size as illustrated by the vertical dashed lines. (b) The tunnelling spectra taken on the boron cluster free surface at 0.1 K, 1.0 K and 10.0 K.  $\Delta_1$ is the label of the hybridization gap size of the spectra as illustrated by the vertical dashed lines. The spectrum obtained at $0.1\ \mathrm{K}$ shows a clear substructure, where the main peak is split into a series of peaks. The dashed blue spectrum is the one taken at $0.1\ \mathrm{K}$ after thermal broadening. $p_0$, $p_1$, $p_2$, $p_3$, $p_4$ label the resonance peaks ($V_b=50~\mathrm{mV}$, $I=0.5~\mathrm{nA}$, $V_{\mathrm{mod}}=0.5\ \mathrm{mV}$). The spectrum obtained at $100\ \mathrm{mK}$ is fitted with three Gaussian peaks ($p_0$ and $p_4$ are too weak to be fitted reliably) and a background between $-30\mathrm{\ mV}$ and $0\mathrm{\ mV}$. Inset: topographies representative of where the ultra-low temperature spectra were taken.}
\label{ltspectroscopy}
\end{figure}

Tunneling spectra obtained at ultra-low temperatures below $100\ \mathrm{mK}$, shown in Fig.~\ref{ltspectroscopy}, exhibit a more complex structure compared to those obtained at higher temperatures (Fig.~\ref{spectroscopy}): (i) The hybridization gap at $E_F$ develops a flat bottom, which is what would be expected for a fully open gap. Nevertheless, there are also in-gap states which reflect the nontriviality of the SmB$_6$ electronic structure. (ii) On top of the boron clusters, the hybridization gap size $\Delta_2$, as illustrated in Fig.~\ref{ltspectroscopy}(a), about 15 meV, is larger than the hybridization gap size $\Delta_1$ (illustrated in Fig. \ref{ltspectroscopy}(b)) on the boron cluster free surface, as observed already at higher temperatures. (iii) The peak at $-8\ \mathrm{meV}$ in the spectra taken on the clean surface is split into a series of resonances which are labelled as $p_0$ to $p_4$ (Fig. \ref{ltspectroscopy}(b)). This structure is reproducible on the clean surface and exhibits only very small variations as a function of location. 

Here we also show the temperature dependence of the tunneling spectra. We can see that at higher temperatures (at 10 K), the resonance peaks are thermally broadened and become consistent with the higher-temperature spectra for surface II shown in the previous section.   

\section{Discussion}
From comparison of the surface topography with the calculations, we deduce that the crystal is cleaved through the boron octahedral layer. We can observe the clean Sm surface and surface with boron clusters. This agrees with the expected difference in work functions of the different possible terminations in the (001) surface, and is also consistent with calculations for other hexaborides \cite{Uijttewaal06}. The observed surface termination in our study is consistent with the fact that pure samarium or boron terminated surfaces are polar. Considering the likely cleavage plane into a layer with boron octahedra, one could expect about half of a layer of boron clusters on the sample surface. In our experiments, the sample is inserted into the cold STM head after cleavage at low temperatures, preventing surface diffusion. We speculate that different surface terminations such as the ($2\times 1$) reconstructions seen in previous STM works may originate from surface reconstruction or diffusion taking place after sample cleavage \cite{yee2013imaging,rossler2014hybridization}. A possible cause is a higher cleavage temperature, enabling activation of surface reconstruction and/or diffusion. In our STM, the cleaving stage is clamped to the 4 K plate of the cryostat, the temperature of which stays well below 20 K during and after cleavage.

The average tunneling spectra of surface I and surface II exhibit a clearly different shape, yet both can be described by a Fano lineshape, which has been widely used to describe tunneling spectra of single impurity systems \cite{ujsaghy_theory_2000,knorr_kondo_2002,wahl_kondo-effect_2009} as well as for Kondo lattices\cite{maltseva_electron_2009,figgins_differential_2010,Schmidt2010, Aynajian2010,Aynajian2012}. Due to the different contributions of direct tunneling to the $f$-orbital and to the conduction band, the resonance can take any shape from a suppression of differential conductance to a peak-like shape. The spectra observed in $\mathrm{SmB}_6$ are quite similar to those observed in other mixed-valent \cite{Wahl2011} or heavy fermion \cite{Schmidt2010, Aynajian2010} compounds. Strong tunneling into the localized $f$-states leads to a peak-like appearance in the tunneling spectra, whereas a reduction of this tunneling channel suppresses the peak and leads to a more asymmetric line shape or even just a gap-like feature. 

Both, surfaces I and II are expected to be polar on a macroscopic scale, and even more so microscopically. It is likely that the boron clusters are negatively charged, as they are in the bulk. The difference in appearance of the boron clusters as well as in local barrier height between adatom free areas of the surface and boron clusters is consistent with our assignment of the termination. For a samarium termination, the largest peak height is hence expected on clean areas, whereas detecting the peak through adsorbates should lead to its reduction. The observed shift in the resonant state towards negative bias voltages on top of clusters suggests a local doping effect due to the surface boron clusters. Qualitatively similar shifts of features in tunneling spectra have been observed in semiconductors and applied to detect the local distribution of electrical charge at the surface of the sample \cite{Dong02}. A similar shift of the resonances has recently been reported from measurements at 300 mK \cite{jiao_additional_2016}.

In the SmB$_6$ crystal, the Sm $4f$ orbital degeneracy is lifted by spin-orbit coupling resulting in $j=5/2$ and $j=7/2$ multiplets that are further split by a crystal electric field (CEF) \cite{kang2015}. The relative position of the $4f$ orbital below the Fermi energy $E_\mathrm{F}$ has been measured by ARPES \cite{frantzeskakis2013kondo, zhu2013polarity}. However, due to the limited energy resolution of the measurements, the CEF effect to the $j=5/2$ bands has not been resolved. $4f$ bands with binding energy from 8 to 20 meV were reported at low temperatures \cite{frantzeskakis2013kondo, miyazaki2012momentum}. A previous STM study \cite{yee2013imaging} has attributed the resonance peak in the conductance spectrum located at $-8$ meV to a $4f$ band close to $E_\mathrm{F}$. Our measurements performed at temperatures well below 1 K reveal that the peak at $-8\ \mathrm{mV}$ really consists of a series of peaks as shown in Fig.~\ref{ltspectroscopy}. The origin of this fine structure can be caused by a few possible reasons. The most plausible scenario is the crystal-field splitting of samarium states in the near surface region: as the surface termination was identified as a samarium termination, the top samarium layer is exposed to a substantially modified environment with lower symmetry than in the bulk, also details of the Kondo screening likely differ from the bulk \cite{alexandrov2015kondo}. A crystal field splitting is expected to lead to satellite peaks of the main Kondo resonance. The topological surface states would be expected to lead to broader spectra features due to their dispersion. However, due to the limited knowledge of the surface properties, e.g., lattice relaxation, magnetic structure, we cannot exclude other mechanisms beyond of Kondo coupling for the observed resonance peaks at low temperature. Different origins such as coupling to low energy bosonic modes of the materials may also cause similar observation.  To fully clarify their origin, more systematic investigation of the surface properties are needed.

On increasing the temperature, the peaks $p_0$ to $p_4$ are broadened and their intensity becomes weaker. The effect is stronger than what would be expected from pure thermal broadening, indicating a Kondo-like mechanism at the origin of the formation of these peaks. However we notice that the full width at half maximum (FWHM) of the resonance peaks (especially for $p_2$ and $p_3$, where it can be determined best) are much smaller than what is expected from the bulk Kondo temperature. The gap minimum remains about the same at $100\ \mathrm{mK}$ with little variation as a function of temperature up to $10\ \mathrm K$.

The in-gap states are well consistent with the predicted nontrivial protected surface states. However, the topological nature of the in-gap states still need to be further confirmed. The inhomogeneity of the surface termination has been discussed with numerical calculations e.g., in Ref. \onlinecite{kim2014termination}. Different surface terminations can shift the surface bands. The variation in gap size could be due to the change of the tunneling matrix due to the local charging effect of the boron cluster. 

\section{Conclusion}
In summary, we have investigated the Kondo hybridization on (001) surfaces of SmB$_6$ with ultra-low temperature STM. With low temperature cleavage, we obtain a surface termination without any reconstruction. By taking spectra at ultra-low temperatures, we clearly confirm the opening of a Kondo hybridization gap and the Kondo gap size of $\sim 8\ \mathrm{meV}$. Even with the fully opened Kondo gap at ultra-low temperature, there is clearly finite density of states within the hybridization gap. This is consistent with the presence of in-gap surface states. We find clear evidence for local electronic effects by the surface boron clusters, which is reflected in shifts of the $4f$ resonant states. Our observations provide new evidence for further decoding of the electronic structures of SmB$_6$.

\section{acknowledgments}
Z.S. gratefully acknowledges X. Dai, P. Liljeroth, H. Yamase, A. Damascelli, G. A. Sawatzky and Y. F. Yang for insightful discussions. V. Duppel performed energy-dispersive x-ray spectroscopy analysis of the sample. Z.S. acknowledges financial support under a Rubicon grant No. 680.50.1119 (NWO, NL). D.S.I. acknowledges support from the German Research Foundation (DFG) under grant No. IN 209/3-2 and the Collaborative Research Center SFB 1143 (project C03).  A.M. and P.W. acknowledge financial support from EPSRC (EP/I031014/1). W.S.P. acknowledges CAPES Foundation, Ministry of Education of Brazil, under grant BEX 9476/13-0. J.J.P. acknowledges the Ministerio de Economia y Competitividad of Spain for financial support under grant No. FIS2016-80434-P, the European Union Seventh Framework Programme under grant agreement No. 604391 Graphene Flagship, the European Union structural funds and the Comunidad de Madrid under grants MAD2D No. S2013/MIT-3007 and NANOFRONTMAG No. S2013/MIT-2850 and the Fundaci$\mathrm{\acute{o}}$n Ram$\mathrm{\acute{o}}$n Areces. W.S.P. and J.J.P. also acknowledge the computer resources and assistance provided by the Centro de Computaci$\mathrm{\acute{o}}$n Cient\'{\i}afica of the Universidad Aut$\mathrm{\acute{o}}$noma de Madrid and the RES. Underpinning data will be made available at \url{http://dx.doi.org/10.17630/33845ed9-4789-4341-86df-2f46d982d7d1}.

\bibliographystyle{apsrev4-1}

\begin{thebibliography}{62}%
\makeatletter
\providecommand \@ifxundefined [1]{%
 \@ifx{#1\undefined}
}%
\providecommand \@ifnum [1]{%
 \ifnum #1\expandafter \@firstoftwo
 \else \expandafter \@secondoftwo
 \fi
}%
\providecommand \@ifx [1]{%
 \ifx #1\expandafter \@firstoftwo
 \else \expandafter \@secondoftwo
 \fi
}%
\providecommand \natexlab [1]{#1}%
\providecommand \enquote  [1]{``#1''}%
\providecommand \bibnamefont  [1]{#1}%
\providecommand \bibfnamefont [1]{#1}%
\providecommand \citenamefont [1]{#1}%
\providecommand \href@noop [0]{\@secondoftwo}%
\providecommand \href [0]{\begingroup \@sanitize@url \@href}%
\providecommand \@href[1]{\@@startlink{#1}\@@href}%
\providecommand \@@href[1]{\endgroup#1\@@endlink}%
\providecommand \@sanitize@url [0]{\catcode `\\12\catcode `\$12\catcode
  `\&12\catcode `\#12\catcode `\^12\catcode `\_12\catcode `\%12\relax}%
\providecommand \@@startlink[1]{}%
\providecommand \@@endlink[0]{}%
\providecommand \url  [0]{\begingroup\@sanitize@url \@url }%
\providecommand \@url [1]{\endgroup\@href {#1}{\urlprefix }}%
\providecommand \urlprefix  [0]{URL }%
\providecommand \Eprint [0]{\href }%
\providecommand \doibase [0]{http://dx.doi.org/}%
\providecommand \selectlanguage [0]{\@gobble}%
\providecommand \bibinfo  [0]{\@secondoftwo}%
\providecommand \bibfield  [0]{\@secondoftwo}%
\providecommand \translation [1]{[#1]}%
\providecommand \BibitemOpen [0]{}%
\providecommand \bibitemStop [0]{}%
\providecommand \bibitemNoStop [0]{.\EOS\space}%
\providecommand \EOS [0]{\spacefactor3000\relax}%
\providecommand \BibitemShut  [1]{\csname bibitem#1\endcsname}%
\let\auto@bib@innerbib\@empty
\bibitem [{\citenamefont {Menth}\ \emph {et~al.}(1969)\citenamefont {Menth},
  \citenamefont {Buehler},\ and\ \citenamefont {Geballe}}]{menth1969magnetic}%
  \BibitemOpen
  \bibfield  {author} {\bibinfo {author} {\bibfnamefont {A.}~\bibnamefont
  {Menth}}, \bibinfo {author} {\bibfnamefont {E.}~\bibnamefont {Buehler}}, \
  and\ \bibinfo {author} {\bibfnamefont {T.~H.}\ \bibnamefont {Geballe}},\
  }\href {\doibase 10.1103/PhysRevLett.22.295} {\bibfield  {journal} {\bibinfo
  {journal} {Phys. Rev. Lett.}\ }\textbf {\bibinfo {volume} {22}},\ \bibinfo
  {pages} {295} (\bibinfo {year} {1969})}\BibitemShut {NoStop}%
\bibitem [{\citenamefont {Nickerson}\ \emph {et~al.}(1971)\citenamefont
  {Nickerson}, \citenamefont {White}, \citenamefont {Lee}, \citenamefont
  {Bachmann}, \citenamefont {Geballe},\ and\ \citenamefont
  {Hull}}]{nickerson1971physical}%
  \BibitemOpen
  \bibfield  {author} {\bibinfo {author} {\bibfnamefont {J.~C.}\ \bibnamefont
  {Nickerson}}, \bibinfo {author} {\bibfnamefont {R.~M.}\ \bibnamefont
  {White}}, \bibinfo {author} {\bibfnamefont {K.~N.}\ \bibnamefont {Lee}},
  \bibinfo {author} {\bibfnamefont {R.}~\bibnamefont {Bachmann}}, \bibinfo
  {author} {\bibfnamefont {T.~H.}\ \bibnamefont {Geballe}}, \ and\ \bibinfo
  {author} {\bibfnamefont {G.~W.}\ \bibnamefont {Hull}},\ }\href {\doibase
  10.1103/PhysRevB.3.2030} {\bibfield  {journal} {\bibinfo  {journal} {Phys.
  Rev. B}\ }\textbf {\bibinfo {volume} {3}},\ \bibinfo {pages} {2030} (\bibinfo
  {year} {1971})}\BibitemShut {NoStop}%
\bibitem [{\citenamefont {Vainshtein}\ \emph {et~al.}(1965)\citenamefont
  {Vainshtein}, \citenamefont {Blokhin},\ and\ \citenamefont
  {Paderno}}]{vainshtein1965}%
  \BibitemOpen
  \bibfield  {author} {\bibinfo {author} {\bibfnamefont {E.}~\bibnamefont
  {Vainshtein}}, \bibinfo {author} {\bibfnamefont {S.}~\bibnamefont {Blokhin}},
  \ and\ \bibinfo {author} {\bibfnamefont {Y.}~\bibnamefont {Paderno}},\
  }\href@noop {} {\bibfield  {journal} {\bibinfo  {journal} {Sov. Phys. Solid
  State}\ }\textbf {\bibinfo {volume} {6}},\ \bibinfo {pages} {2318} (\bibinfo
  {year} {1965})}\BibitemShut {NoStop}%
\bibitem [{\citenamefont {Cohen}\ \emph {et~al.}(1970)\citenamefont {Cohen},
  \citenamefont {Eibsch\"utz},\ and\ \citenamefont
  {West}}]{cohen1970electronic}%
  \BibitemOpen
  \bibfield  {author} {\bibinfo {author} {\bibfnamefont {R.~L.}\ \bibnamefont
  {Cohen}}, \bibinfo {author} {\bibfnamefont {M.}~\bibnamefont {Eibsch\"utz}},
  \ and\ \bibinfo {author} {\bibfnamefont {K.~W.}\ \bibnamefont {West}},\
  }\href {\doibase 10.1103/PhysRevLett.24.383} {\bibfield  {journal} {\bibinfo
  {journal} {Phys. Rev. Lett.}\ }\textbf {\bibinfo {volume} {24}},\ \bibinfo
  {pages} {383} (\bibinfo {year} {1970})}\BibitemShut {NoStop}%
\bibitem [{\citenamefont {Chazalviel}\ \emph {et~al.}(1976)\citenamefont
  {Chazalviel}, \citenamefont {Campagna}, \citenamefont {Wertheim},\ and\
  \citenamefont {Schmidt}}]{chazalviel1976study}%
  \BibitemOpen
  \bibfield  {author} {\bibinfo {author} {\bibfnamefont {J.~N.}\ \bibnamefont
  {Chazalviel}}, \bibinfo {author} {\bibfnamefont {M.}~\bibnamefont
  {Campagna}}, \bibinfo {author} {\bibfnamefont {G.~K.}\ \bibnamefont
  {Wertheim}}, \ and\ \bibinfo {author} {\bibfnamefont {P.~H.}\ \bibnamefont
  {Schmidt}},\ }\href {\doibase 10.1103/PhysRevB.14.4586} {\bibfield  {journal}
  {\bibinfo  {journal} {Phys. Rev. B}\ }\textbf {\bibinfo {volume} {14}},\
  \bibinfo {pages} {4586} (\bibinfo {year} {1976})}\BibitemShut {NoStop}%
\bibitem [{\citenamefont {Frantzeskakis}\ \emph {et~al.}(2013)\citenamefont
  {Frantzeskakis}, \citenamefont {de~Jong}, \citenamefont {Zwartsenberg},
  \citenamefont {Huang}, \citenamefont {Pan}, \citenamefont {Zhang},
  \citenamefont {Zhang}, \citenamefont {Zhang}, \citenamefont {Bao},
  \citenamefont {Tegus}, \citenamefont {Varykhalov}, \citenamefont
  {de~Visser},\ and\ \citenamefont {Golden}}]{frantzeskakis2013kondo}%
  \BibitemOpen
  \bibfield  {author} {\bibinfo {author} {\bibfnamefont {E.}~\bibnamefont
  {Frantzeskakis}}, \bibinfo {author} {\bibfnamefont {N.}~\bibnamefont
  {de~Jong}}, \bibinfo {author} {\bibfnamefont {B.}~\bibnamefont
  {Zwartsenberg}}, \bibinfo {author} {\bibfnamefont {Y.~K.}\ \bibnamefont
  {Huang}}, \bibinfo {author} {\bibfnamefont {Y.}~\bibnamefont {Pan}}, \bibinfo
  {author} {\bibfnamefont {X.}~\bibnamefont {Zhang}}, \bibinfo {author}
  {\bibfnamefont {J.~X.}\ \bibnamefont {Zhang}}, \bibinfo {author}
  {\bibfnamefont {F.~X.}\ \bibnamefont {Zhang}}, \bibinfo {author}
  {\bibfnamefont {L.~H.}\ \bibnamefont {Bao}}, \bibinfo {author} {\bibfnamefont
  {O.}~\bibnamefont {Tegus}}, \bibinfo {author} {\bibfnamefont
  {A.}~\bibnamefont {Varykhalov}}, \bibinfo {author} {\bibfnamefont
  {A.}~\bibnamefont {de~Visser}}, \ and\ \bibinfo {author} {\bibfnamefont
  {M.~S.}\ \bibnamefont {Golden}},\ }\href {\doibase 10.1103/PhysRevX.3.041024}
  {\bibfield  {journal} {\bibinfo  {journal} {Phys. Rev. X}\ }\textbf {\bibinfo
  {volume} {3}},\ \bibinfo {pages} {041024} (\bibinfo {year}
  {2013})}\BibitemShut {NoStop}%
\bibitem [{\citenamefont {Riseborough}(2000)}]{Riseborough2000}%
  \BibitemOpen
  \bibfield  {author} {\bibinfo {author} {\bibfnamefont {P.~S.}\ \bibnamefont
  {Riseborough}},\ }\href {\doibase 10.1080/000187300243345} {\bibfield
  {journal} {\bibinfo  {journal} {Adv. Phys.}\ }\textbf {\bibinfo {volume}
  {49}},\ \bibinfo {pages} {257} (\bibinfo {year} {2000})}\BibitemShut
  {NoStop}%
\bibitem [{\citenamefont {Nanba}\ \emph {et~al.}(1993)\citenamefont {Nanba},
  \citenamefont {Ohta}, \citenamefont {Motokawa}, \citenamefont {Kimura},
  \citenamefont {Kunii},\ and\ \citenamefont {Kasuya}}]{nanba1993gap}%
  \BibitemOpen
  \bibfield  {author} {\bibinfo {author} {\bibfnamefont {T.}~\bibnamefont
  {Nanba}}, \bibinfo {author} {\bibfnamefont {H.}~\bibnamefont {Ohta}},
  \bibinfo {author} {\bibfnamefont {M.}~\bibnamefont {Motokawa}}, \bibinfo
  {author} {\bibfnamefont {S.}~\bibnamefont {Kimura}}, \bibinfo {author}
  {\bibfnamefont {S.}~\bibnamefont {Kunii}}, \ and\ \bibinfo {author}
  {\bibfnamefont {T.}~\bibnamefont {Kasuya}},\ }\href {\doibase
  10.1016/0921-4526(93)90598-Z} {\bibfield  {journal} {\bibinfo  {journal}
  {Physica B: Condensed Matter}\ }\textbf {\bibinfo {volume} {186--188}},\
  \bibinfo {pages} {440} (\bibinfo {year} {1993})}\BibitemShut {NoStop}%
\bibitem [{\citenamefont {Takigawa}\ \emph {et~al.}(1981)\citenamefont
  {Takigawa}, \citenamefont {Yasuoka}, \citenamefont {Kitaoka}, \citenamefont
  {Tanaka}, \citenamefont {Nozaki},\ and\ \citenamefont
  {Ishizawa}}]{takigawa1981nmr}%
  \BibitemOpen
  \bibfield  {author} {\bibinfo {author} {\bibfnamefont {M.}~\bibnamefont
  {Takigawa}}, \bibinfo {author} {\bibfnamefont {H.}~\bibnamefont {Yasuoka}},
  \bibinfo {author} {\bibfnamefont {Y.}~\bibnamefont {Kitaoka}}, \bibinfo
  {author} {\bibfnamefont {T.}~\bibnamefont {Tanaka}}, \bibinfo {author}
  {\bibfnamefont {H.}~\bibnamefont {Nozaki}}, \ and\ \bibinfo {author}
  {\bibfnamefont {Y.}~\bibnamefont {Ishizawa}},\ }\href {\doibase
  10.1143/JPSJ.50.2525} {\bibfield  {journal} {\bibinfo  {journal} {J. Phys.
  Soc. Jpn.}\ }\textbf {\bibinfo {volume} {50}},\ \bibinfo {pages} {2525}
  (\bibinfo {year} {1981})}\BibitemShut {NoStop}%
\bibitem [{\citenamefont {Zhang}\ \emph {et~al.}(2013)\citenamefont {Zhang},
  \citenamefont {Butch}, \citenamefont {Syers}, \citenamefont {Ziemak},
  \citenamefont {Greene},\ and\ \citenamefont
  {Paglione}}]{zhang2013hybridization}%
  \BibitemOpen
  \bibfield  {author} {\bibinfo {author} {\bibfnamefont {X.}~\bibnamefont
  {Zhang}}, \bibinfo {author} {\bibfnamefont {N.~P.}\ \bibnamefont {Butch}},
  \bibinfo {author} {\bibfnamefont {P.}~\bibnamefont {Syers}}, \bibinfo
  {author} {\bibfnamefont {S.}~\bibnamefont {Ziemak}}, \bibinfo {author}
  {\bibfnamefont {R.~L.}\ \bibnamefont {Greene}}, \ and\ \bibinfo {author}
  {\bibfnamefont {J.}~\bibnamefont {Paglione}},\ }\href {\doibase
  10.1103/PhysRevX.3.011011} {\bibfield  {journal} {\bibinfo  {journal} {Phys.
  Rev. X}\ }\textbf {\bibinfo {volume} {3}},\ \bibinfo {pages} {011011}
  (\bibinfo {year} {2013})}\BibitemShut {NoStop}%
\bibitem [{\citenamefont {Gorshunov}\ \emph {et~al.}(1999)\citenamefont
  {Gorshunov}, \citenamefont {Sluchanko}, \citenamefont {Volkov}, \citenamefont
  {Dressel}, \citenamefont {Knebel}, \citenamefont {Loidl},\ and\ \citenamefont
  {Kunii}}]{gorshunov1999low}%
  \BibitemOpen
  \bibfield  {author} {\bibinfo {author} {\bibfnamefont {B.}~\bibnamefont
  {Gorshunov}}, \bibinfo {author} {\bibfnamefont {N.}~\bibnamefont
  {Sluchanko}}, \bibinfo {author} {\bibfnamefont {A.}~\bibnamefont {Volkov}},
  \bibinfo {author} {\bibfnamefont {M.}~\bibnamefont {Dressel}}, \bibinfo
  {author} {\bibfnamefont {G.}~\bibnamefont {Knebel}}, \bibinfo {author}
  {\bibfnamefont {A.}~\bibnamefont {Loidl}}, \ and\ \bibinfo {author}
  {\bibfnamefont {S.}~\bibnamefont {Kunii}},\ }\href {\doibase
  10.1103/PhysRevB.59.1808} {\bibfield  {journal} {\bibinfo  {journal} {Phys.
  Rev. B}\ }\textbf {\bibinfo {volume} {59}},\ \bibinfo {pages} {1808}
  (\bibinfo {year} {1999})}\BibitemShut {NoStop}%
\bibitem [{\citenamefont {Yamaguchi}\ \emph {et~al.}(2013)\citenamefont
  {Yamaguchi}, \citenamefont {Sekiyama}, \citenamefont {Kimura}, \citenamefont
  {Sugiyama}, \citenamefont {Tomida}, \citenamefont {Funabashi}, \citenamefont
  {Komori}, \citenamefont {Balashov}, \citenamefont {Wulfhekel}, \citenamefont
  {Ito}, \citenamefont {Kimura}, \citenamefont {Higashiya}, \citenamefont
  {Tamasaku}, \citenamefont {Yabashi}, \citenamefont {Ishikawa}, \citenamefont
  {Yeo}, \citenamefont {Lee}, \citenamefont {Iga}, \citenamefont {Takabatake},\
  and\ \citenamefont {Suga}}]{yamaguchi2013different}%
  \BibitemOpen
  \bibfield  {author} {\bibinfo {author} {\bibfnamefont {J.}~\bibnamefont
  {Yamaguchi}}, \bibinfo {author} {\bibfnamefont {A.}~\bibnamefont {Sekiyama}},
  \bibinfo {author} {\bibfnamefont {M.~Y.}\ \bibnamefont {Kimura}}, \bibinfo
  {author} {\bibfnamefont {H.}~\bibnamefont {Sugiyama}}, \bibinfo {author}
  {\bibfnamefont {Y.}~\bibnamefont {Tomida}}, \bibinfo {author} {\bibfnamefont
  {G.}~\bibnamefont {Funabashi}}, \bibinfo {author} {\bibfnamefont
  {S.}~\bibnamefont {Komori}}, \bibinfo {author} {\bibfnamefont
  {T.}~\bibnamefont {Balashov}}, \bibinfo {author} {\bibfnamefont
  {W.}~\bibnamefont {Wulfhekel}}, \bibinfo {author} {\bibfnamefont
  {T.}~\bibnamefont {Ito}}, \bibinfo {author} {\bibfnamefont {S.}~\bibnamefont
  {Kimura}}, \bibinfo {author} {\bibfnamefont {A.}~\bibnamefont {Higashiya}},
  \bibinfo {author} {\bibfnamefont {K.}~\bibnamefont {Tamasaku}}, \bibinfo
  {author} {\bibfnamefont {M.}~\bibnamefont {Yabashi}}, \bibinfo {author}
  {\bibfnamefont {T.}~\bibnamefont {Ishikawa}}, \bibinfo {author}
  {\bibfnamefont {S.}~\bibnamefont {Yeo}}, \bibinfo {author} {\bibfnamefont
  {S.-I.}\ \bibnamefont {Lee}}, \bibinfo {author} {\bibfnamefont
  {F.}~\bibnamefont {Iga}}, \bibinfo {author} {\bibfnamefont {T.}~\bibnamefont
  {Takabatake}}, \ and\ \bibinfo {author} {\bibfnamefont {S.}~\bibnamefont
  {Suga}},\ }\href {http://stacks.iop.org/1367-2630/15/i=4/a=043042} {\bibfield
   {journal} {\bibinfo  {journal} {New J. Phys.}\ }\textbf {\bibinfo {volume}
  {15}},\ \bibinfo {pages} {043042} (\bibinfo {year} {2013})}\BibitemShut
  {NoStop}%
\bibitem [{\citenamefont {Gabani}\ \emph {et~al.}(2015)\citenamefont {Gabani},
  \citenamefont {Prist{\'a}{\v{s}}}, \citenamefont {Tak{\'a}{\v{c}}ov{\'a}},
  \citenamefont {Sluchanko}, \citenamefont {Siemensmeyer}, \citenamefont
  {Shitsevalova}, \citenamefont {Filipov},\ and\ \citenamefont
  {Flachbart}}]{gabani2015surface}%
  \BibitemOpen
  \bibfield  {author} {\bibinfo {author} {\bibfnamefont {S.}~\bibnamefont
  {Gabani}}, \bibinfo {author} {\bibfnamefont {G.}~\bibnamefont
  {Prist{\'a}{\v{s}}}}, \bibinfo {author} {\bibfnamefont {I.}~\bibnamefont
  {Tak{\'a}{\v{c}}ov{\'a}}}, \bibinfo {author} {\bibfnamefont {N.}~\bibnamefont
  {Sluchanko}}, \bibinfo {author} {\bibfnamefont {K.}~\bibnamefont
  {Siemensmeyer}}, \bibinfo {author} {\bibfnamefont {N.}~\bibnamefont
  {Shitsevalova}}, \bibinfo {author} {\bibfnamefont {V.}~\bibnamefont
  {Filipov}}, \ and\ \bibinfo {author} {\bibfnamefont {K.}~\bibnamefont
  {Flachbart}},\ }\href {\doibase
  https://doi.org/10.1016/j.solidstatesciences.2015.03.005} {\bibfield
  {journal} {\bibinfo  {journal} {Sol. Stat. Sci.}\ }\textbf {\bibinfo {volume}
  {47}},\ \bibinfo {pages} {17} (\bibinfo {year} {2015})}\BibitemShut {NoStop}%
\bibitem [{\citenamefont {Kasuya}\ \emph {et~al.}(1979)\citenamefont {Kasuya},
  \citenamefont {Takegahara}, \citenamefont {Fujita}, \citenamefont {Tanaka},\
  and\ \citenamefont {Bannai}}]{kasuya1979valence}%
  \BibitemOpen
  \bibfield  {author} {\bibinfo {author} {\bibfnamefont {T.}~\bibnamefont
  {Kasuya}}, \bibinfo {author} {\bibfnamefont {K.}~\bibnamefont {Takegahara}},
  \bibinfo {author} {\bibfnamefont {T.}~\bibnamefont {Fujita}}, \bibinfo
  {author} {\bibfnamefont {T.}~\bibnamefont {Tanaka}}, \ and\ \bibinfo {author}
  {\bibfnamefont {E.}~\bibnamefont {Bannai}},\ }\href
  {https://doi.org/10.1051/jphyscol:19795107} {\bibfield  {journal} {\bibinfo
  {journal} {J. Phys. Coll.}\ }\textbf {\bibinfo {volume} {C5}},\ \bibinfo
  {pages} {308} (\bibinfo {year} {1979})}\BibitemShut {NoStop}%
\bibitem [{\citenamefont {Allen}\ \emph {et~al.}(1979)\citenamefont {Allen},
  \citenamefont {Batlogg},\ and\ \citenamefont {Wachter}}]{allen1979large}%
  \BibitemOpen
  \bibfield  {author} {\bibinfo {author} {\bibfnamefont {J.~W.}\ \bibnamefont
  {Allen}}, \bibinfo {author} {\bibfnamefont {B.}~\bibnamefont {Batlogg}}, \
  and\ \bibinfo {author} {\bibfnamefont {P.}~\bibnamefont {Wachter}},\ }\href
  {\doibase 10.1103/PhysRevB.20.4807} {\bibfield  {journal} {\bibinfo
  {journal} {Phys. Rev. B}\ }\textbf {\bibinfo {volume} {20}},\ \bibinfo
  {pages} {4807} (\bibinfo {year} {1979})}\BibitemShut {NoStop}%
\bibitem [{\citenamefont {Nyhus}\ \emph {et~al.}(1997)\citenamefont {Nyhus},
  \citenamefont {Cooper}, \citenamefont {Fisk},\ and\ \citenamefont
  {Sarrao}}]{nyhus1997low}%
  \BibitemOpen
  \bibfield  {author} {\bibinfo {author} {\bibfnamefont {P.}~\bibnamefont
  {Nyhus}}, \bibinfo {author} {\bibfnamefont {S.~L.}\ \bibnamefont {Cooper}},
  \bibinfo {author} {\bibfnamefont {Z.}~\bibnamefont {Fisk}}, \ and\ \bibinfo
  {author} {\bibfnamefont {J.}~\bibnamefont {Sarrao}},\ }\href {\doibase
  10.1103/PhysRevB.55.12488} {\bibfield  {journal} {\bibinfo  {journal} {Phys.
  Rev. B}\ }\textbf {\bibinfo {volume} {55}},\ \bibinfo {pages} {12488}
  (\bibinfo {year} {1997})}\BibitemShut {NoStop}%
\bibitem [{\citenamefont {Hlawenka}\ \emph {et~al.}(2018)\citenamefont
  {Hlawenka}, \citenamefont {Siemensmeyer}, \citenamefont {Weschke},
  \citenamefont {Varykhalov}, \citenamefont {S{\'a}nchez-Barriga},
  \citenamefont {Shitsevalova}, \citenamefont {Dukhnenko}, \citenamefont
  {Filipov}, \citenamefont {Gab{\'a}ni}, \citenamefont {Flachbart},
  \citenamefont {Rader},\ and\ \citenamefont {Rienks}}]{hlawenka2018samarium}%
  \BibitemOpen
  \bibfield  {author} {\bibinfo {author} {\bibfnamefont {P.}~\bibnamefont
  {Hlawenka}}, \bibinfo {author} {\bibfnamefont {K.}~\bibnamefont
  {Siemensmeyer}}, \bibinfo {author} {\bibfnamefont {E.}~\bibnamefont
  {Weschke}}, \bibinfo {author} {\bibfnamefont {A.}~\bibnamefont {Varykhalov}},
  \bibinfo {author} {\bibfnamefont {J.}~\bibnamefont {S{\'a}nchez-Barriga}},
  \bibinfo {author} {\bibfnamefont {N.}~\bibnamefont {Shitsevalova}}, \bibinfo
  {author} {\bibfnamefont {A.}~\bibnamefont {Dukhnenko}}, \bibinfo {author}
  {\bibfnamefont {V.}~\bibnamefont {Filipov}}, \bibinfo {author} {\bibfnamefont
  {S.}~\bibnamefont {Gab{\'a}ni}}, \bibinfo {author} {\bibfnamefont
  {K.}~\bibnamefont {Flachbart}}, \bibinfo {author} {\bibfnamefont
  {O.}~\bibnamefont {Rader}}, \ and\ \bibinfo {author} {\bibfnamefont
  {E.~D.~L.}\ \bibnamefont {Rienks}},\ }\href {\doibase
  10.1038/s41467-018-02908-7} {\bibfield  {journal} {\bibinfo  {journal} {Nat.
  Commun.}\ }\textbf {\bibinfo {volume} {9}},\ \bibinfo {pages} {517} (\bibinfo
  {year} {2018})}\BibitemShut {NoStop}%
\bibitem [{\citenamefont {Dzero}\ \emph {et~al.}(2010)\citenamefont {Dzero},
  \citenamefont {Sun}, \citenamefont {Galitski},\ and\ \citenamefont
  {Coleman}}]{dzero2010topological}%
  \BibitemOpen
  \bibfield  {author} {\bibinfo {author} {\bibfnamefont {M.}~\bibnamefont
  {Dzero}}, \bibinfo {author} {\bibfnamefont {K.}~\bibnamefont {Sun}}, \bibinfo
  {author} {\bibfnamefont {V.}~\bibnamefont {Galitski}}, \ and\ \bibinfo
  {author} {\bibfnamefont {P.}~\bibnamefont {Coleman}},\ }\href {\doibase
  10.1103/PhysRevLett.104.106408} {\bibfield  {journal} {\bibinfo  {journal}
  {Phys. Rev. Lett.}\ }\textbf {\bibinfo {volume} {104}},\ \bibinfo {pages}
  {106408} (\bibinfo {year} {2010})}\BibitemShut {NoStop}%
\bibitem [{\citenamefont {Dzero}\ \emph {et~al.}(2012)\citenamefont {Dzero},
  \citenamefont {Sun}, \citenamefont {Coleman},\ and\ \citenamefont
  {Galitski}}]{dzero2012theory}%
  \BibitemOpen
  \bibfield  {author} {\bibinfo {author} {\bibfnamefont {M.}~\bibnamefont
  {Dzero}}, \bibinfo {author} {\bibfnamefont {K.}~\bibnamefont {Sun}}, \bibinfo
  {author} {\bibfnamefont {P.}~\bibnamefont {Coleman}}, \ and\ \bibinfo
  {author} {\bibfnamefont {V.}~\bibnamefont {Galitski}},\ }\href {\doibase
  10.1103/PhysRevB.85.045130} {\bibfield  {journal} {\bibinfo  {journal} {Phys.
  Rev. B}\ }\textbf {\bibinfo {volume} {85}},\ \bibinfo {pages} {045130}
  (\bibinfo {year} {2012})}\BibitemShut {NoStop}%
\bibitem [{\citenamefont {Lu}\ \emph {et~al.}(2013)\citenamefont {Lu},
  \citenamefont {Zhao}, \citenamefont {Weng}, \citenamefont {Fang},\ and\
  \citenamefont {Dai}}]{lu2013correlated}%
  \BibitemOpen
  \bibfield  {author} {\bibinfo {author} {\bibfnamefont {F.}~\bibnamefont
  {Lu}}, \bibinfo {author} {\bibfnamefont {J. Z.}~\bibnamefont {Zhao}}, \bibinfo
  {author} {\bibfnamefont {H.}~\bibnamefont {Weng}}, \bibinfo {author}
  {\bibfnamefont {Z.}~\bibnamefont {Fang}}, \ and\ \bibinfo {author}
  {\bibfnamefont {X.}~\bibnamefont {Dai}},\ }\href {\doibase
  10.1103/PhysRevLett.110.096401} {\bibfield  {journal} {\bibinfo  {journal}
  {Phys. Rev. Lett.}\ }\textbf {\bibinfo {volume} {110}},\ \bibinfo {pages}
  {096401} (\bibinfo {year} {2013})}\BibitemShut {NoStop}%
\bibitem [{\citenamefont {Takimoto}(2011)}]{takimoto2011smb6}%
  \BibitemOpen
  \bibfield  {author} {\bibinfo {author} {\bibfnamefont {T.}~\bibnamefont
  {Takimoto}},\ }\href {\doibase 10.1143/JPSJ.80.123710} {\bibfield  {journal}
  {\bibinfo  {journal} {J. Phys. Soc. Jpn.}\ }\textbf {\bibinfo {volume}
  {80}},\ \bibinfo {pages} {123710} (\bibinfo {year} {2011})}\BibitemShut
  {NoStop}%
\bibitem [{\citenamefont {Alexandrov}\ \emph {et~al.}(2013)\citenamefont
  {Alexandrov}, \citenamefont {Dzero},\ and\ \citenamefont
  {Coleman}}]{alexandrov2013cubic}%
  \BibitemOpen
  \bibfield  {author} {\bibinfo {author} {\bibfnamefont {V.}~\bibnamefont
  {Alexandrov}}, \bibinfo {author} {\bibfnamefont {M.}~\bibnamefont {Dzero}}, \
  and\ \bibinfo {author} {\bibfnamefont {P.}~\bibnamefont {Coleman}},\ }\href
  {\doibase 10.1103/PhysRevLett.111.226403} {\bibfield  {journal} {\bibinfo
  {journal} {Phys. Rev. Lett.}\ }\textbf {\bibinfo {volume} {111}},\ \bibinfo
  {pages} {226403} (\bibinfo {year} {2013})}\BibitemShut {NoStop}%
\bibitem [{\citenamefont {Ye}\ \emph {et~al.}()\citenamefont {Ye},
  \citenamefont {Allen},\ and\ \citenamefont {Sun}}]{ye2013topological}%
  \BibitemOpen
  \bibfield  {author} {\bibinfo {author} {\bibfnamefont {M.}~\bibnamefont
  {Ye}}, \bibinfo {author} {\bibfnamefont {J.}~\bibnamefont {Allen}}, \ and\
  \bibinfo {author} {\bibfnamefont {K.}~\bibnamefont {Sun}},\ }\href@noop {}
  {\enquote {\bibinfo {title} {{Topological crystalline Kondo insulators and
  universal topological surface states of SmB$_6$}},}\ }\Eprint
  {http://arxiv.org/abs/arXiv:1307.7191} {arXiv:1307.7191} \BibitemShut
  {NoStop}%
\bibitem [{\citenamefont {Qi}\ and\ \citenamefont
  {Zhang}(2011)}]{RevModPhys.83.1057}%
  \BibitemOpen
  \bibfield  {author} {\bibinfo {author} {\bibfnamefont {X.-L.}\ \bibnamefont
  {Qi}}\ and\ \bibinfo {author} {\bibfnamefont {S.-C.}\ \bibnamefont {Zhang}},\
  }\href {\doibase 10.1103/RevModPhys.83.1057} {\bibfield  {journal} {\bibinfo
  {journal} {Rev. Mod. Phys.}\ }\textbf {\bibinfo {volume} {83}},\ \bibinfo
  {pages} {1057} (\bibinfo {year} {2011})}\BibitemShut {NoStop}%
\bibitem [{\citenamefont {Hasan}\ and\ \citenamefont
  {Kane}(2010)}]{hasan2010colloquium}%
  \BibitemOpen
  \bibfield  {author} {\bibinfo {author} {\bibfnamefont {M.~Z.}\ \bibnamefont
  {Hasan}}\ and\ \bibinfo {author} {\bibfnamefont {C.~L.}\ \bibnamefont
  {Kane}},\ }\href {\doibase 10.1103/RevModPhys.82.3045} {\bibfield  {journal}
  {\bibinfo  {journal} {Rev. Mod. Phys.}\ }\textbf {\bibinfo {volume} {82}},\
  \bibinfo {pages} {3045} (\bibinfo {year} {2010})}\BibitemShut {NoStop}%
\bibitem [{\citenamefont {Neupane}\ \emph {et~al.}(2013)\citenamefont
  {Neupane}, \citenamefont {Alidoust}, \citenamefont {Xu}, \citenamefont
  {Kondo}, \citenamefont {Ishida}, \citenamefont {Kim}, \citenamefont {Liu},
  \citenamefont {Belopolski}, \citenamefont {Jo}, \citenamefont {Chang},
  \citenamefont {Jeng}, \citenamefont {Durakiewicz}, \citenamefont {Balicas},
  \citenamefont {Lin}, \citenamefont {Bansil}, \citenamefont {Shin},
  \citenamefont {Fisk},\ and\ \citenamefont {Hasan}}]{neupane2013surface}%
  \BibitemOpen
  \bibfield  {author} {\bibinfo {author} {\bibfnamefont {M.}~\bibnamefont
  {Neupane}}, \bibinfo {author} {\bibfnamefont {N.}~\bibnamefont {Alidoust}},
  \bibinfo {author} {\bibfnamefont {S.-Y.}\ \bibnamefont {Xu}}, \bibinfo
  {author} {\bibfnamefont {T.}~\bibnamefont {Kondo}}, \bibinfo {author}
  {\bibfnamefont {Y.}~\bibnamefont {Ishida}}, \bibinfo {author} {\bibfnamefont
  {D.}~\bibnamefont {Kim}}, \bibinfo {author} {\bibfnamefont {C.}~\bibnamefont
  {Liu}}, \bibinfo {author} {\bibfnamefont {I.}~\bibnamefont {Belopolski}},
  \bibinfo {author} {\bibfnamefont {Y.}~\bibnamefont {Jo}}, \bibinfo {author}
  {\bibfnamefont {T.-R.}\ \bibnamefont {Chang}}, \bibinfo {author}
  {\bibfnamefont {H.-T.}\ \bibnamefont {Jeng}}, \bibinfo {author}
  {\bibfnamefont {T.}~\bibnamefont {Durakiewicz}}, \bibinfo {author}
  {\bibfnamefont {L.}~\bibnamefont {Balicas}}, \bibinfo {author} {\bibfnamefont
  {H.}~\bibnamefont {Lin}}, \bibinfo {author} {\bibfnamefont {A.}~\bibnamefont
  {Bansil}}, \bibinfo {author} {\bibfnamefont {S.}~\bibnamefont {Shin}},
  \bibinfo {author} {\bibfnamefont {Z.}~\bibnamefont {Fisk}}, \ and\ \bibinfo
  {author} {\bibfnamefont {M.}~\bibnamefont {Hasan}},\ }\href {\doibase
  10.1038/ncomms3991} {\bibfield  {journal} {\bibinfo  {journal} {Nat.
  Commun.}\ }\textbf {\bibinfo {volume} {4}},\ \bibinfo {pages} {2991}
  (\bibinfo {year} {2013})}\BibitemShut {NoStop}%
\bibitem [{\citenamefont {Zhu}\ \emph {et~al.}(2013)\citenamefont {Zhu},
  \citenamefont {Nicolaou}, \citenamefont {Levy}, \citenamefont {Butch},
  \citenamefont {Syers}, \citenamefont {Wang}, \citenamefont {Paglione},
  \citenamefont {Sawatzky}, \citenamefont {Elfimov},\ and\ \citenamefont
  {Damascelli}}]{zhu2013polarity}%
  \BibitemOpen
  \bibfield  {author} {\bibinfo {author} {\bibfnamefont {Z.-H.}\ \bibnamefont
  {Zhu}}, \bibinfo {author} {\bibfnamefont {A.}~\bibnamefont {Nicolaou}},
  \bibinfo {author} {\bibfnamefont {G.}~\bibnamefont {Levy}}, \bibinfo {author}
  {\bibfnamefont {N.~P.}\ \bibnamefont {Butch}}, \bibinfo {author}
  {\bibfnamefont {P.}~\bibnamefont {Syers}}, \bibinfo {author} {\bibfnamefont
  {X.~F.}\ \bibnamefont {Wang}}, \bibinfo {author} {\bibfnamefont
  {J.}~\bibnamefont {Paglione}}, \bibinfo {author} {\bibfnamefont {G.~A.}\
  \bibnamefont {Sawatzky}}, \bibinfo {author} {\bibfnamefont {I.~S.}\
  \bibnamefont {Elfimov}}, \ and\ \bibinfo {author} {\bibfnamefont
  {A.}~\bibnamefont {Damascelli}},\ }\href {\doibase
  10.1103/PhysRevLett.111.216402} {\bibfield  {journal} {\bibinfo  {journal}
  {Phys. Rev. Lett.}\ }\textbf {\bibinfo {volume} {111}},\ \bibinfo {pages}
  {216402} (\bibinfo {year} {2013})}\BibitemShut {NoStop}%
\bibitem [{\citenamefont {Xu}\ \emph {et~al.}(2013)\citenamefont {Xu},
  \citenamefont {Shi}, \citenamefont {Biswas}, \citenamefont {Matt},
  \citenamefont {Dhaka}, \citenamefont {Huang}, \citenamefont {Plumb},
  \citenamefont {Radovi\ifmmode~\acute{c}\else \'{c}\fi{}}, \citenamefont
  {Dil}, \citenamefont {Pomjakushina}, \citenamefont {Conder}, \citenamefont
  {Amato}, \citenamefont {Salman}, \citenamefont {Paul}, \citenamefont {Mesot},
  \citenamefont {Ding},\ and\ \citenamefont {Shi}}]{xu2013surface}%
  \BibitemOpen
  \bibfield  {author} {\bibinfo {author} {\bibfnamefont {N.}~\bibnamefont
  {Xu}}, \bibinfo {author} {\bibfnamefont {X.}~\bibnamefont {Shi}}, \bibinfo
  {author} {\bibfnamefont {P.~K.}\ \bibnamefont {Biswas}}, \bibinfo {author}
  {\bibfnamefont {C.~E.}\ \bibnamefont {Matt}}, \bibinfo {author}
  {\bibfnamefont {R.~S.}\ \bibnamefont {Dhaka}}, \bibinfo {author}
  {\bibfnamefont {Y.}~\bibnamefont {Huang}}, \bibinfo {author} {\bibfnamefont
  {N.~C.}\ \bibnamefont {Plumb}}, \bibinfo {author} {\bibfnamefont
  {M.}~\bibnamefont {Radovi\ifmmode~\acute{c}\else \'{c}\fi{}}}, \bibinfo
  {author} {\bibfnamefont {J.~H.}\ \bibnamefont {Dil}}, \bibinfo {author}
  {\bibfnamefont {E.}~\bibnamefont {Pomjakushina}}, \bibinfo {author}
  {\bibfnamefont {K.}~\bibnamefont {Conder}}, \bibinfo {author} {\bibfnamefont
  {A.}~\bibnamefont {Amato}}, \bibinfo {author} {\bibfnamefont
  {Z.}~\bibnamefont {Salman}}, \bibinfo {author} {\bibfnamefont {D.~M.}\
  \bibnamefont {Paul}}, \bibinfo {author} {\bibfnamefont {J.}~\bibnamefont
  {Mesot}}, \bibinfo {author} {\bibfnamefont {H.}~\bibnamefont {Ding}}, \ and\
  \bibinfo {author} {\bibfnamefont {M.}~\bibnamefont {Shi}},\ }\href {\doibase
  10.1103/PhysRevB.88.121102} {\bibfield  {journal} {\bibinfo  {journal} {Phys.
  Rev. B}\ }\textbf {\bibinfo {volume} {88}},\ \bibinfo {pages} {121102}
  (\bibinfo {year} {2013})}\BibitemShut {NoStop}%
\bibitem [{\citenamefont {Suga}\ \emph {et~al.}(2014)\citenamefont {Suga},
  \citenamefont {Sakamoto}, \citenamefont {Okuda}, \citenamefont {Miyamoto},
  \citenamefont {Kuroda}, \citenamefont {Sekiyama}, \citenamefont {Yamaguchi},
  \citenamefont {Fujiwara}, \citenamefont {Irizawa}, \citenamefont {Ito},
  \citenamefont {Kimura}, \citenamefont {Balashov}, \citenamefont {Wulfhekel},
  \citenamefont {Yeo}, \citenamefont {Iga},\ and\ \citenamefont
  {Imada}}]{suga2014spin}%
  \BibitemOpen
  \bibfield  {author} {\bibinfo {author} {\bibfnamefont {S.}~\bibnamefont
  {Suga}}, \bibinfo {author} {\bibfnamefont {K.}~\bibnamefont {Sakamoto}},
  \bibinfo {author} {\bibfnamefont {T.}~\bibnamefont {Okuda}}, \bibinfo
  {author} {\bibfnamefont {K.}~\bibnamefont {Miyamoto}}, \bibinfo {author}
  {\bibfnamefont {K.}~\bibnamefont {Kuroda}}, \bibinfo {author} {\bibfnamefont
  {A.}~\bibnamefont {Sekiyama}}, \bibinfo {author} {\bibfnamefont
  {J.}~\bibnamefont {Yamaguchi}}, \bibinfo {author} {\bibfnamefont
  {H.}~\bibnamefont {Fujiwara}}, \bibinfo {author} {\bibfnamefont
  {A.}~\bibnamefont {Irizawa}}, \bibinfo {author} {\bibfnamefont
  {T.}~\bibnamefont {Ito}}, \bibinfo {author} {\bibfnamefont {S.}~\bibnamefont
  {Kimura}}, \bibinfo {author} {\bibfnamefont {T.}~\bibnamefont {Balashov}},
  \bibinfo {author} {\bibfnamefont {W.}~\bibnamefont {Wulfhekel}}, \bibinfo
  {author} {\bibfnamefont {S.}~\bibnamefont {Yeo}}, \bibinfo {author}
  {\bibfnamefont {F.}~\bibnamefont {Iga}}, \ and\ \bibinfo {author}
  {\bibfnamefont {S.}~\bibnamefont {Imada}},\ }\href
  {https://doi.org/10.7566/JPSJ.83.014705} {\bibfield  {journal} {\bibinfo
  {journal} {J. Phys. Soc. Jpn.}\ }\textbf {\bibinfo {volume} {83}},\ \bibinfo
  {pages} {014705} (\bibinfo {year} {2014})}\BibitemShut {NoStop}%
\bibitem [{\citenamefont {Wolgast}\ \emph {et~al.}(2013)\citenamefont
  {Wolgast}, \citenamefont {Kurdak}, \citenamefont {Sun}, \citenamefont
  {Allen}, \citenamefont {Kim},\ and\ \citenamefont {Fisk}}]{wolgast2013low}%
  \BibitemOpen
  \bibfield  {author} {\bibinfo {author} {\bibfnamefont {S.}~\bibnamefont
  {Wolgast}}, \bibinfo {author} {\bibfnamefont {{\c{C}}.}~\bibnamefont
  {Kurdak}}, \bibinfo {author} {\bibfnamefont {K.}~\bibnamefont {Sun}},
  \bibinfo {author} {\bibfnamefont {J.~W.}\ \bibnamefont {Allen}}, \bibinfo
  {author} {\bibfnamefont {D.-J.}\ \bibnamefont {Kim}}, \ and\ \bibinfo
  {author} {\bibfnamefont {Z.}~\bibnamefont {Fisk}},\ }\href {\doibase
  10.1103/PhysRevB.88.180405} {\bibfield  {journal} {\bibinfo  {journal} {Phys.
  Rev. B}\ }\textbf {\bibinfo {volume} {88}},\ \bibinfo {pages} {180405}
  (\bibinfo {year} {2013})}\BibitemShut {NoStop}%
\bibitem [{\citenamefont {Zabolotnyy}\ \emph {et~al.}(2018)\citenamefont
  {Zabolotnyy}, \citenamefont {F\"ursich}, \citenamefont {Green}, \citenamefont
  {Lutz}, \citenamefont {Treiber}, \citenamefont {Min}, \citenamefont
  {Dukhnenko}, \citenamefont {Shitsevalova}, \citenamefont {Filipov},
  \citenamefont {Kang}, \citenamefont {Cho}, \citenamefont {Sutarto},
  \citenamefont {He}, \citenamefont {Reinert}, \citenamefont {Inosov},\ and\
  \citenamefont {Hinkov}}]{zabolotnyy2018chemical}%
  \BibitemOpen
  \bibfield  {author} {\bibinfo {author} {\bibfnamefont {V.~B.}\ \bibnamefont
  {Zabolotnyy}}, \bibinfo {author} {\bibfnamefont {K.}~\bibnamefont
  {F\"ursich}}, \bibinfo {author} {\bibfnamefont {R.~J.}\ \bibnamefont
  {Green}}, \bibinfo {author} {\bibfnamefont {P.}~\bibnamefont {Lutz}},
  \bibinfo {author} {\bibfnamefont {K.}~\bibnamefont {Treiber}}, \bibinfo
  {author} {\bibfnamefont {C.-H.}\ \bibnamefont {Min}}, \bibinfo {author}
  {\bibfnamefont {A.~V.}\ \bibnamefont {Dukhnenko}}, \bibinfo {author}
  {\bibfnamefont {N.~Y.}\ \bibnamefont {Shitsevalova}}, \bibinfo {author}
  {\bibfnamefont {V.~B.}\ \bibnamefont {Filipov}}, \bibinfo {author}
  {\bibfnamefont {B.~Y.}\ \bibnamefont {Kang}}, \bibinfo {author}
  {\bibfnamefont {B.~K.}\ \bibnamefont {Cho}}, \bibinfo {author} {\bibfnamefont
  {R.}~\bibnamefont {Sutarto}}, \bibinfo {author} {\bibfnamefont
  {F.}~\bibnamefont {He}}, \bibinfo {author} {\bibfnamefont {F.}~\bibnamefont
  {Reinert}}, \bibinfo {author} {\bibfnamefont {D.~S.}\ \bibnamefont {Inosov}},
  \ and\ \bibinfo {author} {\bibfnamefont {V.}~\bibnamefont {Hinkov}},\ }\href
  {\doibase 10.1103/PhysRevB.97.205416} {\bibfield  {journal} {\bibinfo
  {journal} {Phys. Rev. B}\ }\textbf {\bibinfo {volume} {97}},\ \bibinfo
  {pages} {205416} (\bibinfo {year} {2018})}\BibitemShut {NoStop}%
\bibitem [{\citenamefont {Yee}\ \emph {et~al.}()\citenamefont {Yee},
  \citenamefont {Y.}, \citenamefont {Soumyanarayanan}, \citenamefont {Kim},
  \citenamefont {Fisk},\ and\ \citenamefont {Hoffman}}]{yee2013imaging}%
  \BibitemOpen
  \bibfield  {author} {\bibinfo {author} {\bibfnamefont {M.}~\bibnamefont
  {Yee}}, \bibinfo {author} {\bibfnamefont {H.}~\bibnamefont {Y.}}, \bibinfo
  {author} {\bibfnamefont {A.}~\bibnamefont {Soumyanarayanan}}, \bibinfo
  {author} {\bibfnamefont {D.-J.}\ \bibnamefont {Kim}}, \bibinfo {author}
  {\bibfnamefont {Z.}~\bibnamefont {Fisk}}, \ and\ \bibinfo {author}
  {\bibfnamefont {J.}~\bibnamefont {Hoffman}},\ }\href@noop {} {\enquote
  {\bibinfo {title} {{Imaging the Kondo Insulating Gap on SmB$_6$}},}\ }\Eprint
  {http://arxiv.org/abs/arXiv:1308.1085} {arXiv:1308.1085} \BibitemShut
  {NoStop}%
\bibitem [{\citenamefont {R{\"o}{\ss}ler}\ \emph {et~al.}(2014)\citenamefont
  {R{\"o}{\ss}ler}, \citenamefont {Jang}, \citenamefont {Kim}, \citenamefont
  {Tjeng}, \citenamefont {Fisk}, \citenamefont {Steglich},\ and\ \citenamefont
  {Wirth}}]{rossler2014hybridization}%
  \BibitemOpen
  \bibfield  {author} {\bibinfo {author} {\bibfnamefont {S.}~\bibnamefont
  {R{\"o}{\ss}ler}}, \bibinfo {author} {\bibfnamefont {T.-H.}\ \bibnamefont
  {Jang}}, \bibinfo {author} {\bibfnamefont {D.-J.}\ \bibnamefont {Kim}},
  \bibinfo {author} {\bibfnamefont {L.}~\bibnamefont {Tjeng}}, \bibinfo
  {author} {\bibfnamefont {Z.}~\bibnamefont {Fisk}}, \bibinfo {author}
  {\bibfnamefont {F.}~\bibnamefont {Steglich}}, \ and\ \bibinfo {author}
  {\bibfnamefont {S.}~\bibnamefont {Wirth}},\ }\href {\doibase
  10.1073/pnas.1402643111} {\bibfield  {journal} {\bibinfo  {journal} {Proc.
  Nat. Acad. Sci.}\ }\textbf {\bibinfo {volume} {111}},\ \bibinfo {pages}
  {4798} (\bibinfo {year} {2014})}\BibitemShut {NoStop}%
\bibitem [{\citenamefont {Ruan}\ \emph {et~al.}(2014)\citenamefont {Ruan},
  \citenamefont {Ye}, \citenamefont {Guo}, \citenamefont {Chen}, \citenamefont
  {Chen}, \citenamefont {Zhang},\ and\ \citenamefont
  {Wang}}]{ruan2014emergence}%
  \BibitemOpen
  \bibfield  {author} {\bibinfo {author} {\bibfnamefont {W.}~\bibnamefont
  {Ruan}}, \bibinfo {author} {\bibfnamefont {C.}~\bibnamefont {Ye}}, \bibinfo
  {author} {\bibfnamefont {M.}~\bibnamefont {Guo}}, \bibinfo {author}
  {\bibfnamefont {F.}~\bibnamefont {Chen}}, \bibinfo {author} {\bibfnamefont
  {X.}~\bibnamefont {Chen}}, \bibinfo {author} {\bibfnamefont {G.-M.}\
  \bibnamefont {Zhang}}, \ and\ \bibinfo {author} {\bibfnamefont
  {Y.}~\bibnamefont {Wang}},\ }\href {\doibase 10.1103/PhysRevLett.112.136401}
  {\bibfield  {journal} {\bibinfo  {journal} {Phys. Rev. Lett.}\ }\textbf
  {\bibinfo {volume} {112}},\ \bibinfo {pages} {136401} (\bibinfo {year}
  {2014})}\BibitemShut {NoStop}%
\bibitem [{\citenamefont {Singh}\ \emph {et~al.}(2013)\citenamefont {Singh},
  \citenamefont {Enayat}, \citenamefont {White},\ and\ \citenamefont
  {Wahl}}]{Singh2013}%
  \BibitemOpen
  \bibfield  {author} {\bibinfo {author} {\bibfnamefont {U.}~\bibnamefont
  {Singh}}, \bibinfo {author} {\bibfnamefont {M.}~\bibnamefont {Enayat}},
  \bibinfo {author} {\bibfnamefont {S.}~\bibnamefont {White}}, \ and\ \bibinfo
  {author} {\bibfnamefont {P.}~\bibnamefont {Wahl}},\ }\href {\doibase
  10.1063/1.4788941} {\bibfield  {journal} {\bibinfo  {journal} {Rev. Sci.
  Instr.}\ }\textbf {\bibinfo {volume} {84}},\ \bibinfo {pages} {013708}
  (\bibinfo {year} {2013})}\BibitemShut {NoStop}%
\bibitem [{\citenamefont {Friemel}\ \emph {et~al.}(2012)\citenamefont
  {Friemel}, \citenamefont {Li}, \citenamefont {Dukhnenko}, \citenamefont
  {Shitsevalova}, \citenamefont {Sluchanko}, \citenamefont {Ivanov},
  \citenamefont {Filipov}, \citenamefont {Keimer},\ and\ \citenamefont
  {Inosov}}]{friemel2012resonant}%
  \BibitemOpen
  \bibfield  {author} {\bibinfo {author} {\bibfnamefont {G.}~\bibnamefont
  {Friemel}}, \bibinfo {author} {\bibfnamefont {Y.}~\bibnamefont {Li}},
  \bibinfo {author} {\bibfnamefont {A.~V.}\ \bibnamefont {Dukhnenko}}, \bibinfo
  {author} {\bibfnamefont {N.~Y.}\ \bibnamefont {Shitsevalova}}, \bibinfo
  {author} {\bibfnamefont {N.~E.}\ \bibnamefont {Sluchanko}}, \bibinfo {author}
  {\bibfnamefont {A.}~\bibnamefont {Ivanov}}, \bibinfo {author} {\bibfnamefont
  {V.}~\bibnamefont {Filipov}}, \bibinfo {author} {\bibfnamefont
  {B.}~\bibnamefont {Keimer}}, \ and\ \bibinfo {author} {\bibfnamefont {D.~S.}\
  \bibnamefont {Inosov}},\ }\href {\doibase 10.1038/ncomms1821} {\bibfield
  {journal} {\bibinfo  {journal} {Nat. Commun.}\ }\textbf {\bibinfo {volume}
  {3}},\ \bibinfo {pages} {830} (\bibinfo {year} {2012})}\BibitemShut {NoStop}%
\bibitem [{\citenamefont {Giannozzi}\ \emph {et~al.}(2009)\citenamefont
  {Giannozzi}, \citenamefont {Baroni}, \citenamefont {Bonini}, \citenamefont
  {Calandra}, \citenamefont {Car}, \citenamefont {Cavazzoni}, \citenamefont
  {{Davide Ceresoli}}, \citenamefont {Chiarotti}, \citenamefont {Cococcioni},
  \citenamefont {Dabo}, \citenamefont {Corso}, \citenamefont {Gironcoli},
  \citenamefont {Fabris}, \citenamefont {Fratesi}, \citenamefont {Gebauer},
  \citenamefont {Gerstmann}, \citenamefont {Gougoussis}, \citenamefont {{Anton
  Kokalj}}, \citenamefont {Lazzeri}, \citenamefont {Martin-Samos},
  \citenamefont {Marzari}, \citenamefont {Mauri}, \citenamefont {Mazzarello},
  \citenamefont {{Stefano Paolini}}, \citenamefont {Pasquarello}, \citenamefont
  {Paulatto}, \citenamefont {Sbraccia}, \citenamefont {Scandolo}, \citenamefont
  {Sclauzero}, \citenamefont {Seitsonen}, \citenamefont {Smogunov},
  \citenamefont {Umari},\ and\ \citenamefont
  {Wentzcovitch}}]{giannozzi_quantum_2009}%
  \BibitemOpen
  \bibfield  {author} {\bibinfo {author} {\bibfnamefont {P.}~\bibnamefont
  {Giannozzi}}, \bibinfo {author} {\bibfnamefont {S.}~\bibnamefont {Baroni}},
  \bibinfo {author} {\bibfnamefont {N.}~\bibnamefont {Bonini}}, \bibinfo
  {author} {\bibfnamefont {M.}~\bibnamefont {Calandra}}, \bibinfo {author}
  {\bibfnamefont {R.}~\bibnamefont {Car}}, \bibinfo {author} {\bibfnamefont
  {C.}~\bibnamefont {Cavazzoni}}, \bibinfo {author} {\bibnamefont {{Davide
  Ceresoli}}}, \bibinfo {author} {\bibfnamefont {G.~L.}\ \bibnamefont
  {Chiarotti}}, \bibinfo {author} {\bibfnamefont {M.}~\bibnamefont
  {Cococcioni}}, \bibinfo {author} {\bibfnamefont {I.}~\bibnamefont {Dabo}},
  \bibinfo {author} {\bibfnamefont {A.~D.}\ \bibnamefont {Corso}}, \bibinfo
  {author} {\bibfnamefont {S.~d.}\ \bibnamefont {Gironcoli}}, \bibinfo {author}
  {\bibfnamefont {S.}~\bibnamefont {Fabris}}, \bibinfo {author} {\bibfnamefont
  {G.}~\bibnamefont {Fratesi}}, \bibinfo {author} {\bibfnamefont
  {R.}~\bibnamefont {Gebauer}}, \bibinfo {author} {\bibfnamefont
  {U.}~\bibnamefont {Gerstmann}}, \bibinfo {author} {\bibfnamefont
  {C.}~\bibnamefont {Gougoussis}}, \bibinfo {author} {\bibnamefont {{Anton
  Kokalj}}}, \bibinfo {author} {\bibfnamefont {M.}~\bibnamefont {Lazzeri}},
  \bibinfo {author} {\bibfnamefont {L.}~\bibnamefont {Martin-Samos}}, \bibinfo
  {author} {\bibfnamefont {N.}~\bibnamefont {Marzari}}, \bibinfo {author}
  {\bibfnamefont {F.}~\bibnamefont {Mauri}}, \bibinfo {author} {\bibfnamefont
  {R.}~\bibnamefont {Mazzarello}}, \bibinfo {author} {\bibnamefont {{Stefano
  Paolini}}}, \bibinfo {author} {\bibfnamefont {A.}~\bibnamefont
  {Pasquarello}}, \bibinfo {author} {\bibfnamefont {L.}~\bibnamefont
  {Paulatto}}, \bibinfo {author} {\bibfnamefont {C.}~\bibnamefont {Sbraccia}},
  \bibinfo {author} {\bibfnamefont {S.}~\bibnamefont {Scandolo}}, \bibinfo
  {author} {\bibfnamefont {G.}~\bibnamefont {Sclauzero}}, \bibinfo {author}
  {\bibfnamefont {A.~P.}\ \bibnamefont {Seitsonen}}, \bibinfo {author}
  {\bibfnamefont {A.}~\bibnamefont {Smogunov}}, \bibinfo {author}
  {\bibfnamefont {P.}~\bibnamefont {Umari}}, \ and\ \bibinfo {author}
  {\bibfnamefont {R.~M.}\ \bibnamefont {Wentzcovitch}},\ }\href {\doibase
  10.1088/0953-8984/21/39/395502} {\bibfield  {journal} {\bibinfo  {journal}
  {J. Phys: Condens. Matter}\ }\textbf {\bibinfo {volume} {21}},\ \bibinfo
  {pages} {395502} (\bibinfo {year} {2009})}\BibitemShut {NoStop}%
\bibitem [{\citenamefont {Perdew}\ \emph {et~al.}(1996)\citenamefont {Perdew},
  \citenamefont {Burke},\ and\ \citenamefont
  {Ernzerhof}}]{perdew_generalized_1996}%
  \BibitemOpen
  \bibfield  {author} {\bibinfo {author} {\bibfnamefont {J.~P.}\ \bibnamefont
  {Perdew}}, \bibinfo {author} {\bibfnamefont {K.}~\bibnamefont {Burke}}, \
  and\ \bibinfo {author} {\bibfnamefont {M.}~\bibnamefont {Ernzerhof}},\ }\href
  {\doibase 10.1103/PhysRevLett.77.3865} {\bibfield  {journal} {\bibinfo
  {journal} {Phys. Rev. Lett.}\ }\textbf {\bibinfo {volume} {77}},\ \bibinfo
  {pages} {3865} (\bibinfo {year} {1996})}\BibitemShut {NoStop}%
\bibitem [{\citenamefont {Troullier}\ and\ \citenamefont
  {Martins}(1991)}]{troullier_efficient_1991}%
  \BibitemOpen
  \bibfield  {author} {\bibinfo {author} {\bibfnamefont {N.}~\bibnamefont
  {Troullier}}\ and\ \bibinfo {author} {\bibfnamefont {J.~L.}\ \bibnamefont
  {Martins}},\ }\href {\doibase 10.1103/PhysRevB.43.1993} {\bibfield  {journal}
  {\bibinfo  {journal} {Phys. Rev. B}\ }\textbf {\bibinfo {volume} {43}},\
  \bibinfo {pages} {1993} (\bibinfo {year} {1991})}\BibitemShut {NoStop}%
\bibitem [{\citenamefont {Monkhorst}\ and\ \citenamefont
  {Pack}(1976)}]{monkhorst_special_1976}%
  \BibitemOpen
  \bibfield  {author} {\bibinfo {author} {\bibfnamefont {H.~J.}\ \bibnamefont
  {Monkhorst}}\ and\ \bibinfo {author} {\bibfnamefont {J.~D.}\ \bibnamefont
  {Pack}},\ }\href {\doibase 10.1103/PhysRevB.13.5188} {\bibfield  {journal}
  {\bibinfo  {journal} {Phys. Rev. B}\ }\textbf {\bibinfo {volume} {13}},\
  \bibinfo {pages} {5188} (\bibinfo {year} {1976})}\BibitemShut {NoStop}%
\bibitem [{\citenamefont {Tersoff}\ and\ \citenamefont
  {Hamann}(1985)}]{tersoff_theory_1985}%
  \BibitemOpen
  \bibfield  {author} {\bibinfo {author} {\bibfnamefont {J.}~\bibnamefont
  {Tersoff}}\ and\ \bibinfo {author} {\bibfnamefont {D.~R.}\ \bibnamefont
  {Hamann}},\ }\href {\doibase 10.1103/PhysRevB.31.805} {\bibfield  {journal}
  {\bibinfo  {journal} {Phys. Rev. B}\ }\textbf {\bibinfo {volume} {31}},\
  \bibinfo {pages} {805} (\bibinfo {year} {1985})}\BibitemShut {NoStop}%
\bibitem [{\citenamefont {Koitzsch}\ \emph {et~al.}(2016)\citenamefont
  {Koitzsch}, \citenamefont {Heming}, \citenamefont {Knupfer}, \citenamefont
  {B{\"u}chner}, \citenamefont {Portnichenko}, \citenamefont {Dukhnenko},
  \citenamefont {Shitsevalova}, \citenamefont {Filipov}, \citenamefont {Lev},
  \citenamefont {Strocov}, \citenamefont {Ollivier},\ and\ \citenamefont
  {Inosov}}]{koitzsch2016nesting}%
  \BibitemOpen
  \bibfield  {author} {\bibinfo {author} {\bibfnamefont {A.}~\bibnamefont
  {Koitzsch}}, \bibinfo {author} {\bibfnamefont {N.}~\bibnamefont {Heming}},
  \bibinfo {author} {\bibfnamefont {M.}~\bibnamefont {Knupfer}}, \bibinfo
  {author} {\bibfnamefont {B.}~\bibnamefont {B{\"u}chner}}, \bibinfo {author}
  {\bibfnamefont {P.~Y.}\ \bibnamefont {Portnichenko}}, \bibinfo {author}
  {\bibfnamefont {A.~V.}\ \bibnamefont {Dukhnenko}}, \bibinfo {author}
  {\bibfnamefont {N.~Y.}\ \bibnamefont {Shitsevalova}}, \bibinfo {author}
  {\bibfnamefont {V.~B.}\ \bibnamefont {Filipov}}, \bibinfo {author}
  {\bibfnamefont {L.~L.}\ \bibnamefont {Lev}}, \bibinfo {author} {\bibfnamefont
  {V.~N.}\ \bibnamefont {Strocov}}, \bibinfo {author} {\bibfnamefont
  {J.}~\bibnamefont {Ollivier}}, \ and\ \bibinfo {author} {\bibfnamefont
  {D.~S.}\ \bibnamefont {Inosov}},\ }\href {\doibase 10.1038/ncomms10876}
  {\bibfield  {journal} {\bibinfo  {journal} {Nat. Commun.}\ }\textbf {\bibinfo
  {volume} {7}},\ \bibinfo {pages} {10876} (\bibinfo {year}
  {2016})}\BibitemShut {NoStop}%
\bibitem [{\citenamefont {Smoluchowski}(1941)}]{smoluchowski_anisotropy_1941}%
  \BibitemOpen
  \bibfield  {author} {\bibinfo {author} {\bibfnamefont {R.}~\bibnamefont
  {Smoluchowski}},\ }\href {\doibase 10.1103/PhysRev.60.661} {\bibfield
  {journal} {\bibinfo  {journal} {Phys. Rev.}\ }\textbf {\bibinfo {volume}
  {60}},\ \bibinfo {pages} {661} (\bibinfo {year} {1941})}\BibitemShut
  {NoStop}%
\bibitem [{\citenamefont {Deng}\ \emph {et~al.}(2013)\citenamefont {Deng},
  \citenamefont {Haule},\ and\ \citenamefont {Kotliar}}]{deng2013plutonium}%
  \BibitemOpen
  \bibfield  {author} {\bibinfo {author} {\bibfnamefont {X.}~\bibnamefont
  {Deng}}, \bibinfo {author} {\bibfnamefont {K.}~\bibnamefont {Haule}}, \ and\
  \bibinfo {author} {\bibfnamefont {G.}~\bibnamefont {Kotliar}},\ }\href
  {\doibase 10.1103/PhysRevLett.111.176404} {\bibfield  {journal} {\bibinfo
  {journal} {Phys. Rev. Lett.}\ }\textbf {\bibinfo {volume} {111}},\ \bibinfo
  {pages} {176404} (\bibinfo {year} {2013})}\BibitemShut {NoStop}%
\bibitem [{\citenamefont {Denlinger}\ \emph {et~al.}()\citenamefont
  {Denlinger}, \citenamefont {Allen}, \citenamefont {Kang}, \citenamefont
  {Sun}, \citenamefont {Kim}, \citenamefont {Shim}, \citenamefont {Min},
  \citenamefont {Kim},\ and\ \citenamefont {Fisk}}]{denlinger2013temperature}%
  \BibitemOpen
  \bibfield  {author} {\bibinfo {author} {\bibfnamefont {J.}~\bibnamefont
  {Denlinger}}, \bibinfo {author} {\bibfnamefont {J.}~\bibnamefont {Allen}},
  \bibinfo {author} {\bibfnamefont {J.-S.}\ \bibnamefont {Kang}}, \bibinfo
  {author} {\bibfnamefont {K.}~\bibnamefont {Sun}}, \bibinfo {author}
  {\bibfnamefont {J.-W.}\ \bibnamefont {Kim}}, \bibinfo {author} {\bibfnamefont
  {J.}~\bibnamefont {Shim}}, \bibinfo {author} {\bibfnamefont {B.}~\bibnamefont
  {Min}}, \bibinfo {author} {\bibfnamefont {D.-J.}\ \bibnamefont {Kim}}, \ and\
  \bibinfo {author} {\bibfnamefont {Z.}~\bibnamefont {Fisk}},\ }\href@noop {}
  {\enquote {\bibinfo {title} {{Temperature Dependence of Linked Gap and
  Surface State Evolution in the Mixed Valent Topological Insulator
  SmB$_6$}},}\ }\Eprint {http://arxiv.org/abs/arXiv:1312.6637}
  {arXiv:1312.6637} \BibitemShut {NoStop}%
\bibitem [{\citenamefont {Aynajian}\ \emph {et~al.}(2010)\citenamefont
  {Aynajian}, \citenamefont {da~Silva~Neto}, \citenamefont {Parker},
  \citenamefont {Huang}, \citenamefont {Pasupathy}, \citenamefont {Mydosh},\
  and\ \citenamefont {Yazdani}}]{Aynajian2010}%
  \BibitemOpen
  \bibfield  {author} {\bibinfo {author} {\bibfnamefont {P.}~\bibnamefont
  {Aynajian}}, \bibinfo {author} {\bibfnamefont {E.}~\bibnamefont
  {da~Silva~Neto}}, \bibinfo {author} {\bibfnamefont {C.~V.}\ \bibnamefont
  {Parker}}, \bibinfo {author} {\bibfnamefont {Y.}~\bibnamefont {Huang}},
  \bibinfo {author} {\bibfnamefont {A.}~\bibnamefont {Pasupathy}}, \bibinfo
  {author} {\bibfnamefont {J.}~\bibnamefont {Mydosh}}, \ and\ \bibinfo {author}
  {\bibfnamefont {A.}~\bibnamefont {Yazdani}},\ }\href {\doibase
  10.1073/pnas.1005892107} {\bibfield  {journal} {\bibinfo  {journal} {Proc.
  Nat. Acad. Sci.}\ }\textbf {\bibinfo {volume} {107}},\ \bibinfo {pages}
  {10383} (\bibinfo {year} {2010})}\BibitemShut {NoStop}%
\bibitem [{\citenamefont {Aynajian}\ \emph {et~al.}(2012)\citenamefont
  {Aynajian}, \citenamefont {da~Silva~Neto}, \citenamefont {Gyenis},
  \citenamefont {Baumbach}, \citenamefont {Thompson}, \citenamefont {Fisk},
  \citenamefont {Bauer},\ and\ \citenamefont {Yazdani}}]{Aynajian2012}%
  \BibitemOpen
  \bibfield  {author} {\bibinfo {author} {\bibfnamefont {P.}~\bibnamefont
  {Aynajian}}, \bibinfo {author} {\bibfnamefont {E.~H.}\ \bibnamefont
  {da~Silva~Neto}}, \bibinfo {author} {\bibfnamefont {A.}~\bibnamefont
  {Gyenis}}, \bibinfo {author} {\bibfnamefont {R.~E.}\ \bibnamefont
  {Baumbach}}, \bibinfo {author} {\bibfnamefont {J.~D.}\ \bibnamefont
  {Thompson}}, \bibinfo {author} {\bibfnamefont {Z.}~\bibnamefont {Fisk}},
  \bibinfo {author} {\bibfnamefont {E.~D.}\ \bibnamefont {Bauer}}, \ and\
  \bibinfo {author} {\bibfnamefont {A.}~\bibnamefont {Yazdani}},\ }\href
  {http://dx.doi.org/10.1038/nature11204} {\bibfield  {journal} {\bibinfo
  {journal} {Nature}\ }\textbf {\bibinfo {volume} {486}},\ \bibinfo {pages}
  {201} (\bibinfo {year} {2012})}\BibitemShut {NoStop}%
\bibitem [{\citenamefont {Park}\ \emph {et~al.}(2012)\citenamefont {Park},
  \citenamefont {Tobash}, \citenamefont {Ronning}, \citenamefont {Bauer},
  \citenamefont {Sarrao}, \citenamefont {Thompson},\ and\ \citenamefont
  {Greene}}]{Park2012}%
  \BibitemOpen
  \bibfield  {author} {\bibinfo {author} {\bibfnamefont {W.~K.}\ \bibnamefont
  {Park}}, \bibinfo {author} {\bibfnamefont {P.~H.}\ \bibnamefont {Tobash}},
  \bibinfo {author} {\bibfnamefont {F.}~\bibnamefont {Ronning}}, \bibinfo
  {author} {\bibfnamefont {E.~D.}\ \bibnamefont {Bauer}}, \bibinfo {author}
  {\bibfnamefont {J.~L.}\ \bibnamefont {Sarrao}}, \bibinfo {author}
  {\bibfnamefont {J.~D.}\ \bibnamefont {Thompson}}, \ and\ \bibinfo {author}
  {\bibfnamefont {L.~H.}\ \bibnamefont {Greene}},\ }\href {\doibase
  10.1103/PhysRevLett.108.246403} {\bibfield  {journal} {\bibinfo  {journal}
  {Phys. Rev. Lett.}\ }\textbf {\bibinfo {volume} {108}},\ \bibinfo {pages}
  {246403} (\bibinfo {year} {2012})}\BibitemShut {NoStop}%
\bibitem [{\citenamefont {Uijttewaal}\ \emph {et~al.}(2006)\citenamefont
  {Uijttewaal}, \citenamefont {de~Wijs},\ and\ \citenamefont
  {de~Groot}}]{Uijttewaal06}%
  \BibitemOpen
  \bibfield  {author} {\bibinfo {author} {\bibfnamefont {M.~A.}\ \bibnamefont
  {Uijttewaal}}, \bibinfo {author} {\bibfnamefont {G.~A.}\ \bibnamefont
  {de~Wijs}}, \ and\ \bibinfo {author} {\bibfnamefont {R.~A.}\ \bibnamefont
  {de~Groot}},\ }\href {\doibase 10.1021/jp063347i} {\bibfield  {journal}
  {\bibinfo  {journal} {J. Phys. Chem. B}\ }\textbf {\bibinfo {volume} {110}},\
  \bibinfo {pages} {18459} (\bibinfo {year} {2006})}\BibitemShut {NoStop}%
\bibitem [{\citenamefont {\'Ujs\'aghy}\ \emph {et~al.}(2000)\citenamefont
  {\'Ujs\'aghy}, \citenamefont {Kroha}, \citenamefont {Szunyogh},\ and\
  \citenamefont {Zawadowski}}]{ujsaghy_theory_2000}%
  \BibitemOpen
  \bibfield  {author} {\bibinfo {author} {\bibfnamefont {O.}~\bibnamefont
  {\'Ujs\'aghy}}, \bibinfo {author} {\bibfnamefont {J.}~\bibnamefont {Kroha}},
  \bibinfo {author} {\bibfnamefont {L.}~\bibnamefont {Szunyogh}}, \ and\
  \bibinfo {author} {\bibfnamefont {A.}~\bibnamefont {Zawadowski}},\ }\href
  {\doibase 10.1103/PhysRevLett.85.2557} {\bibfield  {journal} {\bibinfo
  {journal} {Phys. Rev. Lett.}\ }\textbf {\bibinfo {volume} {85}},\ \bibinfo
  {pages} {2557} (\bibinfo {year} {2000})}\BibitemShut {NoStop}%
\bibitem [{\citenamefont {Knorr}\ \emph {et~al.}(2002)\citenamefont {Knorr},
  \citenamefont {Schneider}, \citenamefont {Diekh\"oner}, \citenamefont
  {Wahl},\ and\ \citenamefont {Kern}}]{knorr_kondo_2002}%
  \BibitemOpen
  \bibfield  {author} {\bibinfo {author} {\bibfnamefont {N.}~\bibnamefont
  {Knorr}}, \bibinfo {author} {\bibfnamefont {M.~A.}\ \bibnamefont
  {Schneider}}, \bibinfo {author} {\bibfnamefont {L.}~\bibnamefont
  {Diekh\"oner}}, \bibinfo {author} {\bibfnamefont {P.}~\bibnamefont {Wahl}}, \
  and\ \bibinfo {author} {\bibfnamefont {K.}~\bibnamefont {Kern}},\ }\href
  {\doibase 10.1103/PhysRevLett.88.096804} {\bibfield  {journal} {\bibinfo
  {journal} {Phys. Rev. Lett.}\ }\textbf {\bibinfo {volume} {88}},\ \bibinfo
  {pages} {096804} (\bibinfo {year} {2002})}\BibitemShut {NoStop}%
\bibitem [{\citenamefont {Wahl}\ \emph {et~al.}(2009)\citenamefont {Wahl},
  \citenamefont {Seitsonen}, \citenamefont {Diekhöner}, \citenamefont
  {Schneider},\ and\ \citenamefont {Kern}}]{wahl_kondo-effect_2009}%
  \BibitemOpen
  \bibfield  {author} {\bibinfo {author} {\bibfnamefont {P.}~\bibnamefont
  {Wahl}}, \bibinfo {author} {\bibfnamefont {A.~P.}\ \bibnamefont {Seitsonen}},
  \bibinfo {author} {\bibfnamefont {L.}~\bibnamefont {Diekhöner}}, \bibinfo
  {author} {\bibfnamefont {M.~A.}\ \bibnamefont {Schneider}}, \ and\ \bibinfo
  {author} {\bibfnamefont {K.}~\bibnamefont {Kern}},\ }\href {\doibase
  10.1088/1367-2630/11/11/113015} {\bibfield  {journal} {\bibinfo  {journal}
  {New J. Phys.}\ }\textbf {\bibinfo {volume} {11}},\ \bibinfo {pages} {113015}
  (\bibinfo {year} {2009})}\BibitemShut {NoStop}%
\bibitem [{\citenamefont {Maltseva}\ \emph {et~al.}(2009)\citenamefont
  {Maltseva}, \citenamefont {Dzero},\ and\ \citenamefont
  {Coleman}}]{maltseva_electron_2009}%
  \BibitemOpen
  \bibfield  {author} {\bibinfo {author} {\bibfnamefont {M.}~\bibnamefont
  {Maltseva}}, \bibinfo {author} {\bibfnamefont {M.}~\bibnamefont {Dzero}}, \
  and\ \bibinfo {author} {\bibfnamefont {P.}~\bibnamefont {Coleman}},\ }\href
  {\doibase 10.1103/PhysRevLett.103.206402} {\bibfield  {journal} {\bibinfo
  {journal} {Phys. Rev. Lett.}\ }\textbf {\bibinfo {volume} {103}},\ \bibinfo
  {pages} {206402} (\bibinfo {year} {2009})}\BibitemShut {NoStop}%
\bibitem [{\citenamefont {Figgins}\ and\ \citenamefont
  {Morr}(2010)}]{figgins_differential_2010}%
  \BibitemOpen
  \bibfield  {author} {\bibinfo {author} {\bibfnamefont {J.}~\bibnamefont
  {Figgins}}\ and\ \bibinfo {author} {\bibfnamefont {D.~K.}\ \bibnamefont
  {Morr}},\ }\href {\doibase 10.1103/PhysRevLett.104.187202} {\bibfield
  {journal} {\bibinfo  {journal} {Phys. Rev. Lett.}\ }\textbf {\bibinfo
  {volume} {104}},\ \bibinfo {pages} {187202} (\bibinfo {year}
  {2010})}\BibitemShut {NoStop}%
\bibitem [{\citenamefont {Schmidt}\ \emph {et~al.}(2010)\citenamefont
  {Schmidt}, \citenamefont {Hamidian}, \citenamefont {Wahl}, \citenamefont
  {Meier}, \citenamefont {Balatsky}, \citenamefont {Garrett}, \citenamefont
  {J.Williams}, \citenamefont {Luke},\ and\ \citenamefont
  {Davis}}]{Schmidt2010}%
  \BibitemOpen
  \bibfield  {author} {\bibinfo {author} {\bibfnamefont {A.}~\bibnamefont
  {Schmidt}}, \bibinfo {author} {\bibfnamefont {M.~H.}\ \bibnamefont
  {Hamidian}}, \bibinfo {author} {\bibfnamefont {P.}~\bibnamefont {Wahl}},
  \bibinfo {author} {\bibfnamefont {F.}~\bibnamefont {Meier}}, \bibinfo
  {author} {\bibfnamefont {A.~V.}\ \bibnamefont {Balatsky}}, \bibinfo {author}
  {\bibfnamefont {J.~D.}\ \bibnamefont {Garrett}}, \bibinfo {author}
  {\bibfnamefont {T.}~\bibnamefont {J.Williams}}, \bibinfo {author}
  {\bibfnamefont {G.~M.}\ \bibnamefont {Luke}}, \ and\ \bibinfo {author}
  {\bibfnamefont {J.~C.}\ \bibnamefont {Davis}},\ }\href {\doibase
  10.1038/nature09073} {\bibfield  {journal} {\bibinfo  {journal} {Nature}\
  }\textbf {\bibinfo {volume} {465}},\ \bibinfo {pages} {570} (\bibinfo {year}
  {2010})}\BibitemShut {NoStop}%
\bibitem [{\citenamefont {Wahl}\ \emph {et~al.}(2011)\citenamefont {Wahl},
  \citenamefont {Diekh\"oner}, \citenamefont {Schneider}, \citenamefont
  {Treubel}, \citenamefont {Lin},\ and\ \citenamefont {Kern}}]{Wahl2011}%
  \BibitemOpen
  \bibfield  {author} {\bibinfo {author} {\bibfnamefont {P.}~\bibnamefont
  {Wahl}}, \bibinfo {author} {\bibfnamefont {L.}~\bibnamefont {Diekh\"oner}},
  \bibinfo {author} {\bibfnamefont {M.~A.}\ \bibnamefont {Schneider}}, \bibinfo
  {author} {\bibfnamefont {F.}~\bibnamefont {Treubel}}, \bibinfo {author}
  {\bibfnamefont {C.~T.}\ \bibnamefont {Lin}}, \ and\ \bibinfo {author}
  {\bibfnamefont {K.}~\bibnamefont {Kern}},\ }\href {\doibase
  10.1103/PhysRevB.84.245131} {\bibfield  {journal} {\bibinfo  {journal} {Phys.
  Rev. B}\ }\textbf {\bibinfo {volume} {84}},\ \bibinfo {pages} {245131}
  (\bibinfo {year} {2011})}\BibitemShut {NoStop}%
\bibitem [{\citenamefont {Y.~Dong}\ and\ \citenamefont {Ploog}(2002)}]{Dong02}%
  \BibitemOpen
  \bibfield  {author} {\bibinfo {author} {\bibfnamefont {R.~H.}\ \bibnamefont
  {Y.~Dong}, \bibfnamefont {R.~M.~Feenstra}}\ and\ \bibinfo {author}
  {\bibfnamefont {K.~H.}\ \bibnamefont {Ploog}},\ }\href
  {https://doi.org/10.1116/1.1491535} {\bibfield  {journal} {\bibinfo
  {journal} {J. Vac. Sci. Tech. B}\ }\textbf {\bibinfo {volume} {20}},\
  \bibinfo {pages} {1677} (\bibinfo {year} {2002})}\BibitemShut {NoStop}%
\bibitem [{\citenamefont {Jiao}\ \emph {et~al.}(2016)\citenamefont {Jiao},
  \citenamefont {R{\"o}{\ss}ler}, \citenamefont {Kim}, \citenamefont {Tjeng},
  \citenamefont {Fisk}, \citenamefont {Steglich},\ and\ \citenamefont
  {Wirth}}]{jiao_additional_2016}%
  \BibitemOpen
  \bibfield  {author} {\bibinfo {author} {\bibfnamefont {L.}~\bibnamefont
  {Jiao}}, \bibinfo {author} {\bibfnamefont {S.}~\bibnamefont
  {R{\"o}{\ss}ler}}, \bibinfo {author} {\bibfnamefont {D.}~\bibnamefont {Kim}},
  \bibinfo {author} {\bibfnamefont {L.}~\bibnamefont {Tjeng}}, \bibinfo
  {author} {\bibfnamefont {Z.}~\bibnamefont {Fisk}}, \bibinfo {author}
  {\bibfnamefont {F.}~\bibnamefont {Steglich}}, \ and\ \bibinfo {author}
  {\bibfnamefont {S.}~\bibnamefont {Wirth}},\ }\href {\doibase
  10.1038/ncomms13762} {\bibfield  {journal} {\bibinfo  {journal} {Nat.
  Commun.}\ }\textbf {\bibinfo {volume} {7}},\ \bibinfo {pages} {13762}
  (\bibinfo {year} {2016})}\BibitemShut {NoStop}%
\bibitem [{\citenamefont {Kang}\ \emph {et~al.}(2015)\citenamefont {Kang},
  \citenamefont {Kim}, \citenamefont {Kim}, \citenamefont {Kang}, \citenamefont
  {Denlinger},\ and\ \citenamefont {Min}}]{kang2015}%
  \BibitemOpen
  \bibfield  {author} {\bibinfo {author} {\bibfnamefont {C.-J.}\ \bibnamefont
  {Kang}}, \bibinfo {author} {\bibfnamefont {J.}~\bibnamefont {Kim}}, \bibinfo
  {author} {\bibfnamefont {K.}~\bibnamefont {Kim}}, \bibinfo {author}
  {\bibfnamefont {J.}~\bibnamefont {Kang}}, \bibinfo {author} {\bibfnamefont
  {J.}~\bibnamefont {Denlinger}}, \ and\ \bibinfo {author} {\bibfnamefont
  {B.}~\bibnamefont {Min}},\ }\href {\doibase 10.7566/JPSJ.84.024722}
  {\bibfield  {journal} {\bibinfo  {journal} {J. Phys. Soc. Jpn.}\ }\textbf
  {\bibinfo {volume} {84}},\ \bibinfo {pages} {024722} (\bibinfo {year}
  {2015})}\BibitemShut {NoStop}%
\bibitem [{\citenamefont {Miyazaki}\ \emph {et~al.}(2012)\citenamefont
  {Miyazaki}, \citenamefont {Hajiri}, \citenamefont {Ito}, \citenamefont
  {Kunii},\ and\ \citenamefont {Kimura}}]{miyazaki2012momentum}%
  \BibitemOpen
  \bibfield  {author} {\bibinfo {author} {\bibfnamefont {H.}~\bibnamefont
  {Miyazaki}}, \bibinfo {author} {\bibfnamefont {T.}~\bibnamefont {Hajiri}},
  \bibinfo {author} {\bibfnamefont {T.}~\bibnamefont {Ito}}, \bibinfo {author}
  {\bibfnamefont {S.}~\bibnamefont {Kunii}}, \ and\ \bibinfo {author}
  {\bibfnamefont {S. I.}\ \bibnamefont {Kimura}},\ }\href {\doibase
  10.1103/PhysRevB.86.075105} {\bibfield  {journal} {\bibinfo  {journal} {Phys.
  Rev. B}\ }\textbf {\bibinfo {volume} {86}},\ \bibinfo {pages} {075105}
  (\bibinfo {year} {2012})}\BibitemShut {NoStop}%
\bibitem [{\citenamefont {Alexandrov}\ \emph {et~al.}(2015)\citenamefont
  {Alexandrov}, \citenamefont {Coleman},\ and\ \citenamefont
  {Erten}}]{alexandrov2015kondo}%
  \BibitemOpen
  \bibfield  {author} {\bibinfo {author} {\bibfnamefont {V.}~\bibnamefont
  {Alexandrov}}, \bibinfo {author} {\bibfnamefont {P.}~\bibnamefont {Coleman}},
  \ and\ \bibinfo {author} {\bibfnamefont {O.}~\bibnamefont {Erten}},\ }\href
  {\doibase 10.1103/PhysRevLett.114.177202} {\bibfield  {journal} {\bibinfo
  {journal} {Phys. Rev. Lett.}\ }\textbf {\bibinfo {volume} {114}},\ \bibinfo
  {pages} {177202} (\bibinfo {year} {2015})}\BibitemShut {NoStop}%
\bibitem [{\citenamefont {Kim}\ \emph {et~al.}(2014)\citenamefont {Kim},
  \citenamefont {Kim}, \citenamefont {Kang}, \citenamefont {Kim}, \citenamefont
  {Choi}, \citenamefont {Kang}, \citenamefont {Denlinger},\ and\ \citenamefont
  {Min}}]{kim2014termination}%
  \BibitemOpen
  \bibfield  {author} {\bibinfo {author} {\bibfnamefont {J.}~\bibnamefont
  {Kim}}, \bibinfo {author} {\bibfnamefont {K.}~\bibnamefont {Kim}}, \bibinfo
  {author} {\bibfnamefont {C.-J.}\ \bibnamefont {Kang}}, \bibinfo {author}
  {\bibfnamefont {S.}~\bibnamefont {Kim}}, \bibinfo {author} {\bibfnamefont
  {H.~C.}\ \bibnamefont {Choi}}, \bibinfo {author} {\bibfnamefont {J.-S.}\
  \bibnamefont {Kang}}, \bibinfo {author} {\bibfnamefont {J.~D.}\ \bibnamefont
  {Denlinger}}, \ and\ \bibinfo {author} {\bibfnamefont {B.~I.}\ \bibnamefont
  {Min}},\ }\href {\doibase 10.1103/PhysRevB.90.075131} {\bibfield  {journal}
  {\bibinfo  {journal} {Phys. Rev. B}\ }\textbf {\bibinfo {volume} {90}},\
  \bibinfo {pages} {075131} (\bibinfo {year} {2014})}\BibitemShut {NoStop}%
\end{thebibliography}

\end{document}